\documentclass[useAMS,usenatbib]{mn2e}%\documentclass{mn2e}
\usepackage{epsfig}
\usepackage{epsf}
\usepackage{graphicx}
\usepackage{natbib}
\usepackage{amsmath}\usepackage{amssymb}\usepackage{stfloats}

\title[The Nature of LBVs]{Luminous Blue Variables are Antisocial:
  Their Isolation Implies they are Kicked Mass Gainers in Binary
  Evolution}
\author[Smith \& Tombleson]{Nathan Smith$^1$\thanks{Email:
    nathans@as.arizona.edu} \& Ryan Tombleson$^1$ \\ $^1$Steward
  Observatory, University of Arizona, 933 North Cherry Avenue, Tucson,
  AZ 85721, USA}

\begin{document}
\date{Accepted 0000, Received 0000, in original form 0000}
\pagerange{\pageref{firstpage}--\pageref{lastpage}} \pubyear{2002}
\def\arcdeg{\degr}
\maketitle
\label{firstpage}

\begin{abstract}

  Based on their relatively isolated environments, we argue that
  luminous blue variables (LBVs) must be primarily the product of
  binary evolution, challenging the traditional single-star view
  wherein LBVs mark a brief transition between massive O-type stars
  and Wolf-Rayet (WR) stars.  If the latter were true, then LBVs
  should be concentrated in young massive clusters like early O-type
  stars.  This is decidedly not the case.  Examining locations of LBVs
  in our Galaxy and the Magellanic Clouds reveals that, with only a
  few exceptions, LBVs systematically avoid clusters of O-type
  stars. In the Large Magellanic Cloud, LBVs are statistically much
  more isolated than O-type stars, and (perhaps most surprisingly)
  even more isolated than WR stars.  This makes it impossible for LBVs
  to be single ``massive stars in transition'' to WR stars.  Instead,
  we propose that massive stars and supernova (SN) subtypes are
  dominated by bifurcated evolutionary paths in interacting binaries,
  wherein most WR stars and SNe~Ibc correspond to the mass donors,
  while LBVs (and their lower-mass analogs like B[e] supergiants,
  which are even more isolated) are the mass gainers.  In this view,
  LBVs are evolved massive blue stragglers.  Through binary mass
  transfer, rejuvinated mass gainers get enriched, spun up, and
  sometimes kicked far from their clustered birthsites by their
  companion's SN.  This scenario agrees better with LBVs exploding as
  Type~IIn SNe in isolation, and it predicts that many massive runaway
  stars may be rapid rotators.
  % We argue that environmental trends of various SN subtypes are
  % influenced more by binarity and SN kicks, rather than tracing
  % initial mass as is generally assumed.
  Mergers or blue Thorne-Zykow-like objects might also give rise to
  LBVs, but these scenarios may have a harder time explaining why LBVs
  avoid clusters.

\end{abstract}


% note - look at these:
% http://adsabs.harvard.edu/abs/2011AAS...21733818E
% http://adsabs.harvard.edu/abs/2011AAS...21715205S

%%%%%%%%%%%%%%%%%%%%%%%%%%%%%%%%%%%%%%%%%%%%%%%%%%%%%%%%%%%%%%%%%%%%%%
\section{INTRODUCTION}

Mass loss is inexorably linked to evolution for high-mass stars.  In
fact, it has a {\it deterministic} influence, which in turn has
tremendous impact on other areas of astronomy influenced by stellar
feedback (regulating star formation, galaxy evolution, chemical
evolution, reionization, etc.).  For most of their lives, massive
stars above $\sim$20 $M_{\odot}$ shed mass in fast winds that affect
their subsequent evolution, but in post-main sequence (post-MS) phases
the mass loss becomes critical in determining the type of resulting
supernova (SN).  The most dramatic mass loss in post-MS evolution is
during the luminous blue variable (LBV) phase.

Some of the most pressing issues in massive star research concern
resolving problems with mass-loss rates and how they are incorporated
into evolutionary models, as well as the treatment (or neglect) of
close binaries in these models (see recent reviews by
\citealt{smith14,langer12}).  A majority of massive stars are in
binary systems whose separation is small enough that they will
exchange mass
\citep{sana12,sana08,sana09,kk12,kiminki12,chip14,chini12,mahy09,
  evans06,debecker06,gm01,gies87}.  With current uncertainties in mass
loss and mass transfer, connecting massive stars to their end fates as
various types of SN explosions is still very challenging.  LBVs and
their eruptive mass loss have emerged as the linchpin in our paradigm
for the evolution of single massive stars.  In light of recent
estimates of lower mass-loss rates for clumpy winds, eruptive mass
loss of LBVs would need to be the primary agent responsible for
turning most H-rich O stars (above $\sim$30-35 $M_{\odot}$, the upper
limit for red supergiants) into H-free Wolf-Rayet (WR) stars before
they die as SNe~Ibc \citep{smithowocki06}. However, our understanding
of the physics and evolutionary states of LBVs is still poor, and
their connection to (or overlap with) the evolution of close binary
systems leaves many open questions.  When ideas about LBVs were taking
shape, \citet{jsg89} discussed mass-transferring binaries and how they
may provide suitable explanations for LBVs, but for whatever reason,
these ideas did not appear to dominate the interpretation of LBVs and
so they were mostly discussed as very massive single-stars.  Later,
\citet{soker04} mentioned difficulties in explaining the formation of
the nebula around $\eta$ Carinae with a single star, although that
paper did not discuss evolutionary aspects of LBVs.  The role of
binarity and initial mass in evolution is a central question we
address here from a new perspective, by studying LBV environments.

LBVs were recognized early-on as the brightest blue irregular
variables in nearby spiral galaxies like M31, M33, and NGC~2403
\citep{hs53,ts68}, originally referred to as the ``Hubble-Sandage
variables''.  Famous Galactic objects like P~Cygni and $\eta$ Carinae
had spectacular outbursts in the 17th and 19th centuries,
respectively, and appeared to share many of the same properties, so
they were grouped together and called ``LBVs'' by \citet{conti84}.
Over a dozen true LBVs are now identified in the Milky Way (MW) and
the Large and Small Magellanic Clouds (LMC/SMC), and a similar number
resides in other Local Group galaxies
\citep{hd94,vg01,svdk04,clark05}.  Stars that spectroscopically
resemble LBVs with similar luminosity and color, but which have not
(as yet) been observed to show the signature eruptive variability of
LBVs, are often called ``LBV candidates''; these are presumed to be
temporarily dormant LBVs.

% here we go

The traditional view of LBVs, which emerged in the 1980s and 1990s, is
that they correspond to a very brief {\it transitional} phase of
evolution of the most massive single stars, when the star moves from
core H burning, through the onset of H shell burning, to the core He
burning phase \citep{hd94,ln02}.  The motivation for a very brief
phase comes from the fact that LBVs are extremely rare compared to O
stars: the duration of the LBV phase is thought to be only a few
10$^4$ yr \citep{hd94} based on this line of reasoning.\footnote{Some
  recent single-star models indicate longer LBV phases of 1-2 $\times$
  10$^5$ yr \citep{groh14}, but it is unclear if this longer LBV
  lifetime (and shorter WR lifetime) agrees with observed statistics.}
A typical and often quoted monotonic evolutionary scheme for a star of
60-100 $M_{\odot}$ initially is:

\smallskip 

\noindent O star $\rightarrow$ Of/WNH $\rightarrow$ LBV $\rightarrow$
WN $\rightarrow$ WC $\rightarrow$ SN Ibc.

\smallskip

\noindent In this scenario, the strong mass-loss experienced by LBVs
is a crucial component of the central paradigm of massive star
evolution, which is that a star's own mass loss determines its
evolution.  High luminosity powers strong radiatively driven winds,
and this wind mass loss propels the evolution that converts the most
massive H-rich main-sequence stars to become H-free WR stars.  This is
the co-called ``Conti scenario'' \citep{conti76}.  LBVs are crucial to
this picture because of the recent downward revisions of O-star
mass-loss rates, as noted above (see review by \citealt{smith14} and
many references therein).  Thus, single massive stars of 30-80
$M_{\odot}$ do not have strong enough winds to make WR stars on their
own; LBV eruptions are needed \citep{smithowocki06}.\footnote{It may
  be possible that near-Eddington luminosities in the most massive O
  stars around 100 $M_{\odot}$ (especially WNH stars), might be strong
  enough to make WR stars without an intervening LBV phase
  \citep{grafener11}.  In this case, the remaining He cores would be
  more massive than any observed H-free WR stars
  \citep{smithconti08,crowther07}.  Such massive He stars might be
  rare, so this is not yet a solved issue.}

%%%%%%%%%%%%%%%

The LBV phase is still generally presumed to arise in the evolution of
single stars \citep{groh14}, and it is {\it required} in the standard
scenario for single massive-star evolution. The origin of the
instability is not fully understood, but it is expected that the core
luminosity goes up with time as the growing He core contracts, while
winds steadily reduce the stellar mass.  Thus, the ratio $L/M$ goes up
until the star confronts the classical Eddington limit and somehow
erupts with catastrophic mass loss \citep{hd94,maeder92,uf98}.  As
seen in the case of $\eta$ Car, these eruptions can eject 10-20
$M_{\odot}$ in a single event \citep{smith03}.

%%%%%%%%%%%%%%%%%%%%%%%%%%%%%%%%%%%%%%%%%%%%%%%%%%%%%%%%%%%%%%%%%%%%%%%
\begin{figure*}\begin{center}
\includegraphics[width=5.9in]{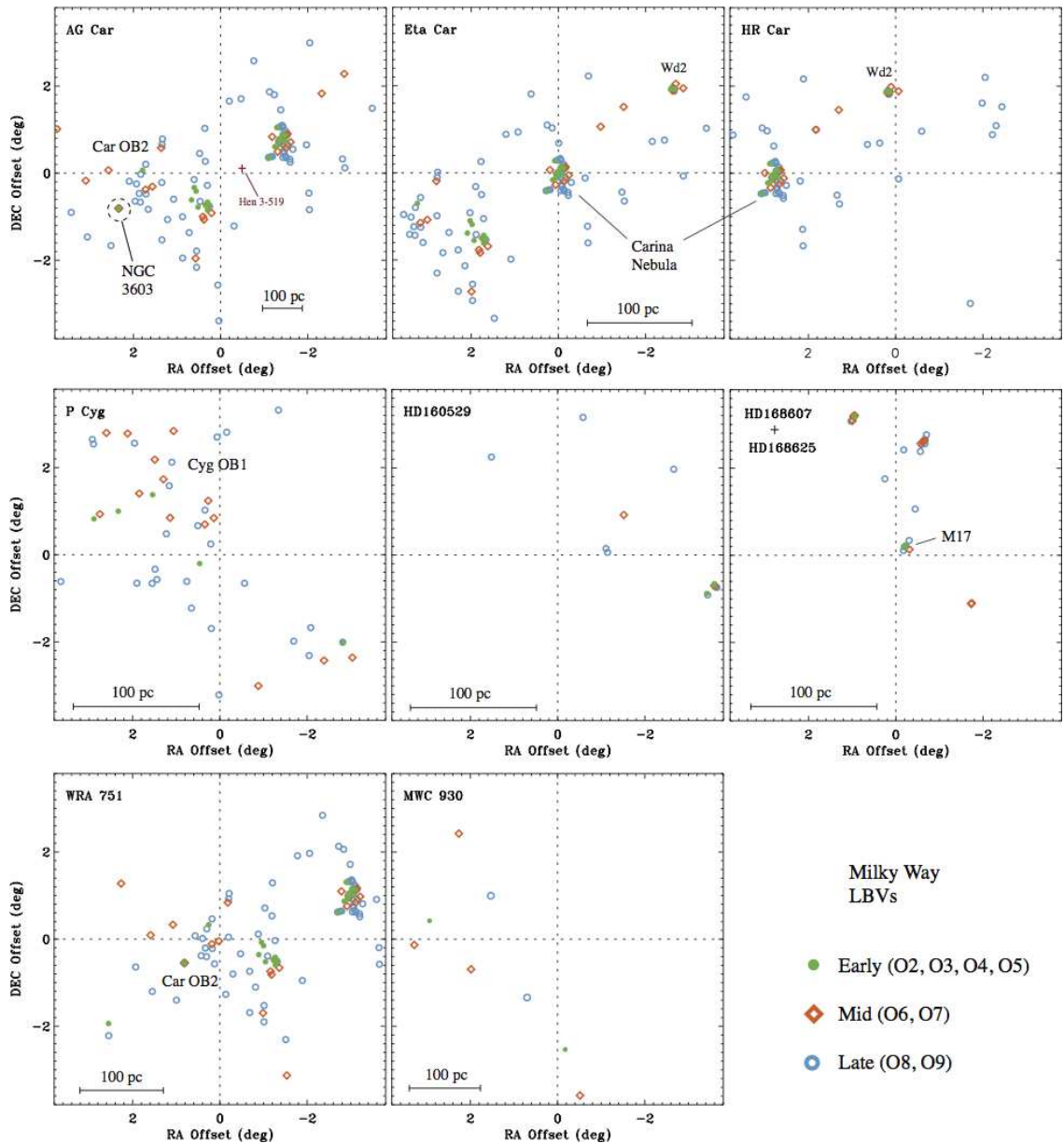}
\end{center}
\caption{Each panel shows the distribution of known O-type stars as
  seen projected on the sky around the LBV listed in the upper left
  corner (that LBV is located at the origin of each plot).  AG Car,
  $\eta$ Car, HR Car, and Wra 751 are all in the same part of the sky
  and their plots overlap, but they are at different distances, so we
  show a separate plot for each.  The location of the LBV candidate
  Hen 3-519 is indicated in the AG Car plot, and the candidate
  HD~168625 and the LBV HD~168607 are at almost the same position,
  outside M17.  Three different bins of O star spectral types are
  color coded: Solid green circles are early O stars (O5 and earlier),
  orange diamonds are mid O stars (O6-O7), and blue unfilled circles
  are late O stars (O8 and O9).  A very rough size scale is noted in
  each panel, appropriate for the (often uncertain) distance to that
  LBV.}
\label{fig:mw}
\end{figure*}
%%%%%%%%%%%%%%%%%%%%%%%%%%%%%%%%%%%%%%%%%%%%%%%%%%%%%%%%%%%%%%%%%%%%%%%

%%%%%%%%%%%%%%%%%%%%%%%%%%%%%%%%%%%%%%%%%%%%%%%%%%%%%%%%%%%%%%%%%%%%%%%
\begin{figure*}\begin{center}
\includegraphics[width=5.9in]{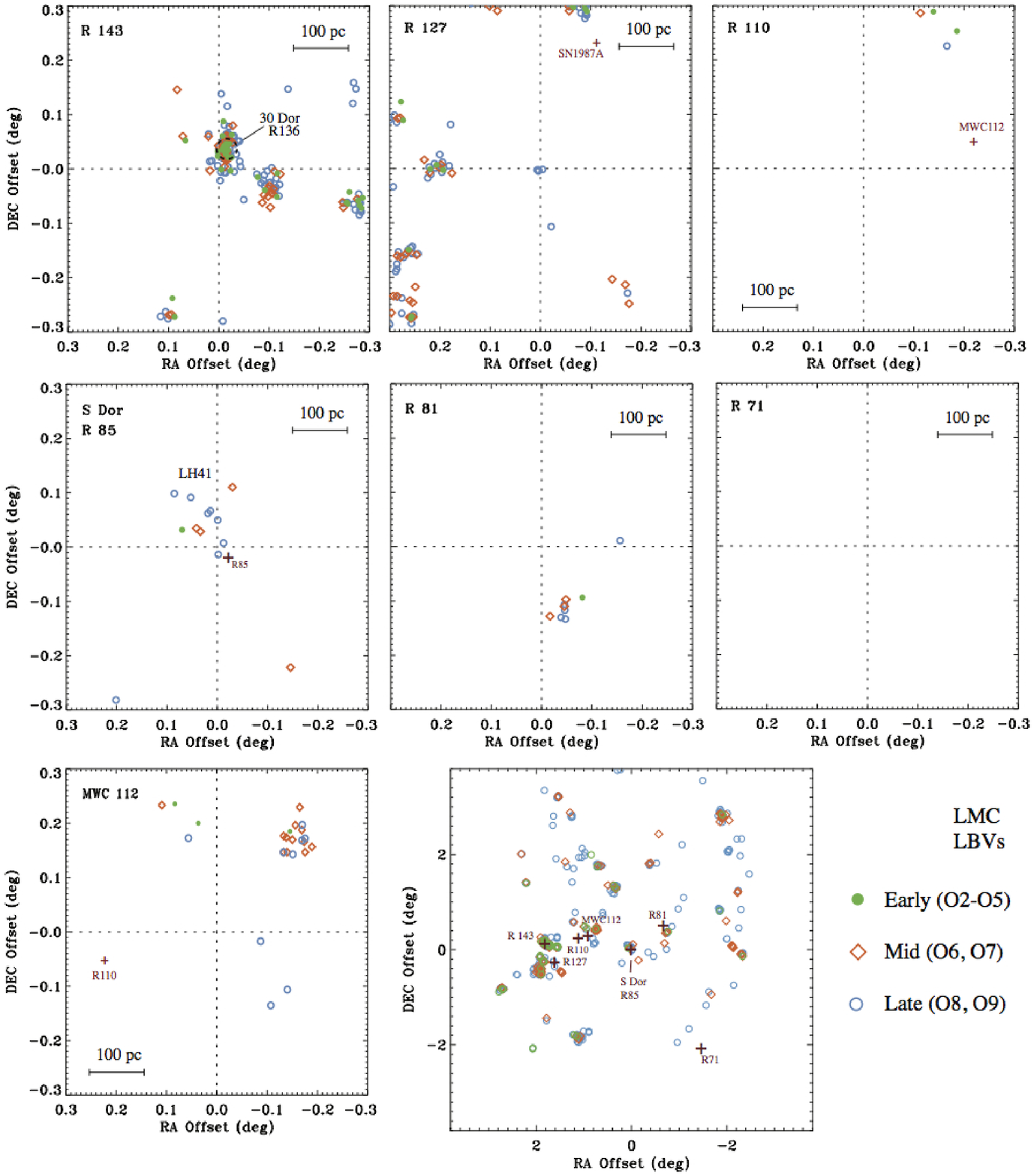}
\end{center}
\caption{Same as Figure~\ref{fig:mw}, but for the confirmed LBVs in
  the LMC. We do not include a separate plot for R85, since it is very
  close to S Dor in the same association.  We also include a panel
  that shows the locations of all LBVs in the LMC for reference.  Note
  that the plot for R~71 is blank (no O-type stars around).}
\label{fig:lmc}
\end{figure*}
%%%%%%%%%%%%%%%%%%%%%%%%%%%%%%%%%%%%%%%%%%%%%%%%%%%%%%%%%%%%%%%%%%%%%%%

%{\bf Some Problems with the Traditional View:} 

While this traditional view of LBVs persisted through the 1990s
\citep{hd94}, in the past decade or so, several problems have emerged
that threaten it.  One issue has to do with the interpretation of LBV
variability and the behavior of their winds.  The proposed
interpretation was that during a standard S~Dor eruption, the star
brightens while staying at constant bolometric luminosity because its
mass-loss rate is assumed to increase, which in turn initiates a
cooler and brighter ``pseudo photosphere'' \citep{hd94}.  However,
quantitative spectroscopy has since revealed that mass-loss rates in
S~Dor outbursts do not increase enough to cause true pseudo
photospheres \citep{dekoter96,groh09}, and the changes do not really
occur at constant bolometric luminosity \citep{groh09}. Instead, the
brightening in an S~Dor event is more akin to a pulsation or temporary
inflation of the envelope, perhaps driven by the sub-surface Fe
opacity peak \citep{grafener12}.

Similarly, the traditional explanation for LBV giant eruptions was
that they experience a substantial {\it increase} in their bolometric
radiative luminosity (for reasons unexplained) that temporarily pushes
them well above the classical Eddington limit \citep{hd94,hds99}; this
necessarily initiates a super-Eddington continuum-driven wind
\citep{owocki04}. However, several lines of evidence now point to an
explosive mechanism driving LBV giant eruptions, instead of (or in
addition to) a wind \citep{smith08,smith13,smith03}.  Moreover,
spectra of the light echoes from $\eta$ Car's 19th century Great
Eruption seem inconsistent with a wind pseudo-photosphere
\citep{rest12,prieto14}.

These considerations point to problems with our interpretation of the
physics behind LBV eruptions, but even more serious problems have
arisen with the traditional role played by LBVs in evolution.  The
first clear sign of something very fishy came from the recognition
that eruptive LBVs are the most likely progenitors of a particular
class of luminous narrow-lined SNe with Type~IIn spectra (SNe~IIn).
The dense H-rich circumstellar material (CSM) of some SNe~IIn requires
eruptive pre-SN mass loss akin to LBVs, and in some cases very large
CSM mass budgets (see review by \citealt{smith14}).  Other aspects of
the CSM around SNe~IIn point to wind variability reminiscent of LBVs
\citep{trundle08,kv06,gv11}.  Moreover, we now have four direct
detections of LBV-like progenitors of SNe~IIn, including SN~1961V,
SN~2005gl, SN~2010jl, and SN~2009ip (\citealt{gl09,smith10,smith11c};
see review by \citealt{smith14}).  In the traditional paradigm of
single-star evolution, massive LBVs are prohibited from exploding as
SNe~IIn, because LBVs represent merely a brief intermediate phase
after core-H burning and before core-He burning.  They should still
have 0.5-1 Myr to live as a WR star, and should die without their H
envelopes.

Another clue that something was seriously amiss came from ``SN
impostors'', which are thought to be LBV giant eruptions associated
with very massive stars in external galaxies \citep{smith11a}.  For
some nearby SN impostors, detections of their relatively
low-luminosity dust-enshrouded progenitors, as well as the ages of
surrounding stars, suggest either progenitors that are 10-15
$M_{\odot}$ BSGs/RSGs or even $\sim$8 $M_{\odot}$ super-AGB stars
(e.g., \citealt{prieto08,thompson09,kochanek11,gogarten09}). This is
much lower than the initial masses attributed to eruptive LBVs, which
are generally above 25 $M_{\odot}$ \citep{svdk04}.  The reason this is
physically significant is because these lower-mass stars ($\sim$8
$M_{\odot}$) never approach the Eddington limit in their normal
evolution, but they appear to suffer LBV-like eruptions anyway.  If
they can do this, then the physical cause of the outbursts might not
be related exclusively to high initial mass, but perhaps some other
exotic mechanism (collisions, mergers, etc.).  Additionally, Anderson
and collaborators \citep{aj09,anderson12,habergham14} examined the
spatial correlation between various types of extragalactic SNe or SN
impostors, as compared to presumed indicators of youth such as
H$\alpha$ emission in their host galaxies.  On the one hand, they
found the expected result that SNe~II-P (arising from 8-20 $M_{\odot}$
stars; \citealt{smartt09}) are less correlated with H$\alpha$ than
stripped-envelope SNe.  However, they also found the very surprising
result that {\it SN impostors and SNe~IIn are apparently even less
  correlated with star formation than SNe II-P}. This was curious, and
motivated us to take another look at LBVs in the MW and LMC/SMC.

%binary models
In this paper, we will argue that the likely solution to many of these
problems with LBVs resides in binary evolution, and that the resulting
SN kicks may be quite important in explaining their isolation.  The
idea that binary interaction influences the outcomes of stellar
evolution is not a new one \citep{paczynski67}, and various aspects
including SN kicks have been studied in theoretical work by several
groups \citep{pod92,vanrensbergen96,dvb98,dvb07,bb99,izzard04,
  cantiello07,demink07,demink13,eldridge08,eldridge11,es09}.
Uncertainty persists in the difficult problems of mass transfer and
the subsequent evolution of the mass gainer.  The way that LBVs may
fit into this picture is disussed in more detail later in the paper.

We begin with a discussion of the projected locations on the sky of
classic Galactic and Magellanic Cloud LBVs as compared to the
locations of O-type stars (\S 2).  This includes a qualitative
examination of the sites of individual LBVs, as well as a quantitative
statistical comparison of the relative isolation of early, mid, and
late-type O stars, WR stars, LBVs, B[e] supergiants, and RSGs.  For
this statistical comparison, we only consider stars in the LMC and
SMC, due to large distance and reddening uncertainties in the Milky
Way.  In \S 3 we discuss how these results are fundamentally
incompatible with the standard monotonic evolutionary view of LBVs as
the descendants of very massive single stars, and in \S 4 we argue for
an alternative view of bifurcated evolutionary paths for massive stars
(where the ``bifurcation'' corresponds to mass donors and mass
gainers).  In \S 5 we discuss the environments of extragalactic SN
impostors and SNe.  We end (\S 6) with an epilogue about why something
so simple as the avoidance of OB-star clusters by LBVs was not
emphasized before.

%%%%%%%%%%%%%%%%%%%%%%%%%%%%%%%%%%%%%%%%%%%%%%%%%%%%%%%%%%%%%%%%%%%%%%%
\begin{figure}\begin{center}
\includegraphics[width=2.8in]{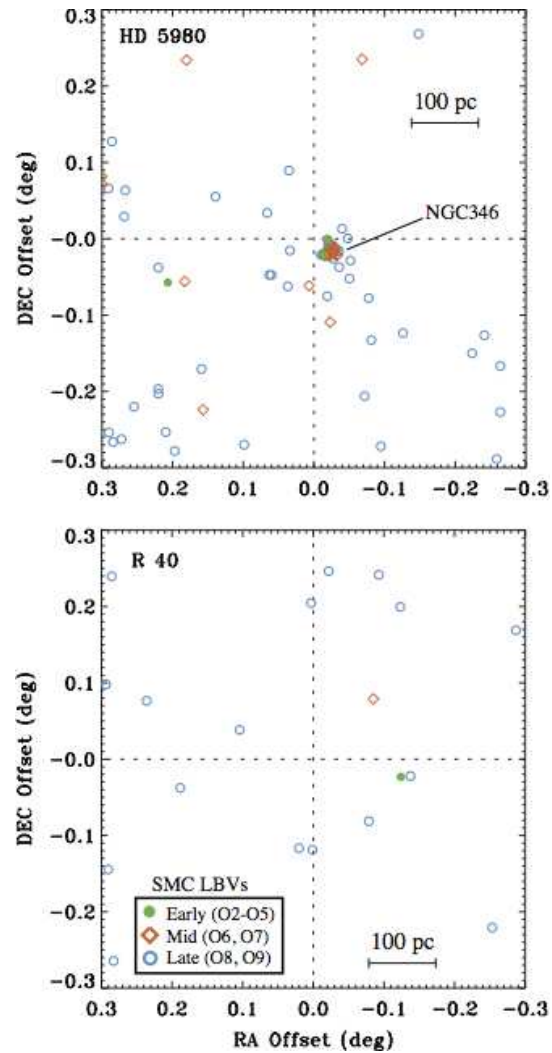}
\end{center}
\caption{Same as Figure~\ref{fig:mw}, but for the two confirmed
  LBVs in the SMC.}
\label{fig:smc}
\end{figure}
%%%%%%%%%%%%%%%%%%%%%%%%%%%%%%%%%%%%%%%%%%%%%%%%%%%%%%%%%%%%%%%%%%%%%%%

\section{POSITIONS OF LBVS RELATIVE TO O-TYPE STARS AND YOUNG
  CLUSTERS}

\subsection{Environments of Individual LBVs}

In Figures~\ref{fig:mw}, \ref{fig:lmc}, and \ref{fig:smc} we present
maps of O-type stars in the surrounding environments for several
individual cases of well-studied LBVs. The positions of O-type stars
are taken from the SIMBAD
databse.\footnote{http://simbad.u-strasbg.fr/simbad/} This depends on
the spectral type listed in SIMBAD, which in some cases contains
errors.  As a check, we also made similar plots using the very
recently revised Galactic O-star Catalog \citep{ma13}.  Using this
newer catalog (not shown), some spectral types changed slightly (for
example, some B0 stars changed to late O-types and vice versa, and
some O stars shifted from our early-type bin to late-type or vice
versa), but this did not change the overall result described below
where LBVs generally avoid clusters of O stars.  It did not, for
example, produce O-star clusters where there were none previously.

For Galactic LBVs we plot a projected size on the sky of $\pm$3.8 deg,
and for stars in the Magellanic Clouds we plot $\pm$0.3 deg.  In each
case this represents (very roughly) the surrounding several 10$^2$ pc
of each LBV.  This is the 2D location projected on the sky, which
gives the LBVs the ``benefit of the doubt'' that they are actually at
the same distance as the O-type stars seen near them on the sky.  We
did not attempt to correct for distances along the line-of-sight to
further cull the sample of O-type stars, because that may introduce a
bias.  For example, in many cases, such as Milky Way LBVs whose
distances are based on radial velocities, the distance adopted might
be substantially wrong if the star has a peculiar velocity.  Since our
main finding is that LBVs are relatively isolated, treating all stars
projected on the sky as being at the same distance is conservative.

%%%%%%%%%%%%%%%%%%%%%%%%%%%%%%%%%%%%%%%%%%%%%%%%%%%%%%%%%%%%%%%%%%%%%%%
\begin{center}
\begin{table*}\begin{minipage}{5.5in}%
      \caption{LBVs and their Nearby O stars in SIMBAD}\scriptsize
\begin{tabular}{@{}llccccc}\hline\hline
LBV        &Galaxy   &$M_{\rm eff}$  &D1  &D2  &ST1  &ST2  \\ %\hline       
(name)     &(name)   &($M_{\odot}$) &(deg)&(deg) & & \\ \hline
%%%
$\eta$ Car &MW       &250    &0.011   &0.016  &O9.5 V &O8.5 V    \\ %\hline       
AG Car     &MW       &100    &0.413   &0.442  &O8     &O9.5      \\ %\hline       
(Hen 3-519)&MW       &45     &0.646   &0.697  &O8     &O9.5      \\ %\hline       
HR Car     &MW       &35     &0.148   &0.784  &O9 II  &O9 Ib     \\ %\hline       
P Cyg      &MW       &55     &0.325   &0.507  &O9 III &O5e       \\ %\hline       
HD160529   &MW       &32     &1.119   &1.145  &O9 III &O8 III    \\ %\hline       
HD168607   &MW       &(30)   &0.195   &0.262  &O8 V   &O5 V      \\ %\hline       
(HD168625) &MW       &28     &0.206   &0.272  &O8 V   &O5 V      \\ %\hline       
Wra 751    &MW       &50     &0.051   &0.195  &O8 V   &O9 V      \\ %\hline       
MWC930     &MW       &32     &1.511   &1.828  &O8.5V  &O8n       \\ \hline       
%%%%
R143       &LMC      &60    &0.00519 &0.00641 &O3.5 III &O8 Ib     \\ %\hline       
R127       &LMC      &90    &0.00475 &0.00547 &O9.7 II  &O8.5 II   \\ %\hline       
S Dor      &LMC      &55    &0.0138  &0.0144  &O8.5 V   &O8.5 V    \\ %\hline       
R81        &LMC      &(40)  &0.1236  &0.1291  &O5 V     &O6 Ib     \\ %\hline       
R110       &LMC      &30    &0.2805  &0.3080  &O8.5 V   &O5 V      \\ %\hline       
R71        &LMC      &29    &0.4448  &0.4714  &O9 II    &O8.5 V    \\ %\hline       
MWC112     &LMC      &(60)  &0.0892  &0.1729  &O9 V     &O8 III    \\ %\hline       
R85        &LMC      &28    &0.0252  &0.0288  &O8.5 V   &O8.5 V    \\ %\hline       
(R84)      &LMC      &30    &0.1575  &0.2276  &O9 II    &O9 II     \\ %\hline       
(R99)      &LMC      &30    &0.0412  &0.0832  &O9 II    &O9.5 V    \\ %\hline       
(R126)     &LMC      &(40)  &0.0358  &0.0836  &O9.5 I   &O8.5 V    \\ %\hline       
(S61)      &LMC      &90    &0.1432  &0.1542  &O7 III   &O9.5 III  \\ %\hline       
(S119)     &LMC      &50    &0.3467  &0.3694  &O8.5 V   &O9 V      \\ %\hline       
(Sk-69142a)&LMC      &60    &0.0522  &0.1524  &O9 V     &O9.5 V    \\ %\hline       
(Sk-69279) &LMC      &52    &0.0685  &0.0993  &O9.5 III &O8 V      \\ %\hline       
(Sk-69271) &LMC      &50    &0.0406  &0.0593  &O9.5 III &O8 V      \\ \hline       
%%%
HD5980     &SMC      &150   &0.0191  &0.0191  &O8 V     &O5.5 V    \\ %\hline       
R40        &SMC      &32    &0.1112  &0.1131  &O8 V     &O9 V      \\ %\hline       
(R4)       &SMC      &(30)  &0.0160  &0.0560  &O8.5 V   &O9.5 V    \\ %\hline       
%%%%
\hline
\end{tabular}
%%%%%%%%%%%%%%%%%%%%%%%%%%%%%%%%%%%%%%%%%%%%%%%%%%%%%%%%%%%%%%%%%%%%%%%%%
%%%%%%%%%%%%%%%%%%%%%%%%%%%%

\medskip

Notes:  \\
Names of LBV candidates are noted in parentheses. \\
$M_{\rm eff}$ = Adopted effective initial ZAMS mass (not a
measurement) based on single-star evolutionary tracks appropriate for the star's present-day luminosity (see \S 4, and Figure~\ref{fig:hrd}). For Galactic objects, uncertainty is dominated by the distance and is hard to quantify.  For MC sources, uncertainties are dominated by assumptions about the treatment of interior physics in the evolutionary models.  Those with relatively poorly constrained luminosity and mass have the representative mass in parentheses. \\
D1 = projected distance to nearest O star (ignoring bound companions); used in Figure~\ref{fig:cum}. \\
D2 = projected distance to second-nearest O star; used in Figure~\ref{fig:cum2}. \\
ST1 = spectral type of nearest O star. \\
ST2 = spectral type of 2nd nearest O star. \\
%%%%%%%%%%%%%%%%%%%%%%%%%%%%%%%%%%%%%%%%%%%%%%%%%%%%%%%%%%%%%%%%%%%%%%%%
\end{minipage}\end{table*}\label{tab:list}
\end{center}

In this study, we concentrate on confirmed LBVs that have relatively
low visual-wavelength extinction, plus a few well-studied and
unobscured LBV candidates with shells.  We therefore exclude LBVs and
candidates in the Galactic Center or Wd1, for example (see
\citealt{clark05,clark09}).  The reason we exclude these is because we
examine the spatial distribution of O-type stars around LBVs, but the
sample and distribution of O-types stars may be highly incomplete in
regions with high visible obscuration.  Moreover, the instances of
known LBVs with high extinction will be biased to those residing in
dense clusters, since their LBV-like nature is often discovered
serendipitously when their host cluster is studied.  Our selection
criteria are therefore that we include all confirmed LBVs in the MW,
LMC, and SMC, except those that are highly reddened.  We also include
LBV candidates with a massive CSM shell that likely indicates a
previous LBV-like giant eruption, but again we exclude those that are
heavily reddened.  The discovery of circumstellar LBV nebulae (usually
accomplished with narrow-band imaging) is not biased against detecting
them in star clusters (e.g., $\eta$ Car and R127 are both in clusters
and their nebulae are easily detected), so it is unlikely that the
inclusion of LBV candidates with shells will bias the results.  There
is no detection bias against recognizing LBVs themselves in clusters,
since any LBV should be the visually brightest member in a cluster.
The potential bias works the other way --- i.e. that the O stars may
be harder to detect next to a bright LBV.  However, it is unlikely
that clusters of O stars surrounding LBVs have escaped detection,
since LBVs have been studied with high-contrast imaging to detect
faint nebular shells (e.g., \citealt{weis03,stahl87}).

The list of specific LBVs and LBV candidates we consider is given in
Table~1, and plots of the spatial distribution of O-type stars
surrounding each confirmed LBV are given in Figure~\ref{fig:mw} for
LBVs in the MW, and Figures~\ref{fig:lmc} and \ref{fig:smc} for the
LMC and SMC, respectively.  Table~1 lists various measured quantities
for these LBVs, including D1, D2, ST1, ST2, and $M_{\rm eff}$.  D1 and
D2 are the projected distances on the sky (in degrees) between the LBV
and the nearest O star (D1) and the second-nearest O-type star (D2).
D1 and D2 do not include companion O stars when an LBV is in a bound
binary or multiple system.  For the LMC and SMC targets, these values
are used for the distributions plotted in Figures~\ref{fig:cum} and
\ref{fig:cum2}, respectively.  ST1 and ST2 are the corresponding
spectral types of these nearest and second-nearest O-type stars, for
reference.  One can see that in most cases, the nearest O stars are
relatively late-type, rather than early O-type stars.  $M_{\rm eff}$
is a rough estimate of the effective initial mass of the star, as a
guide for what to expect if one were to assume that single-star
evolutionary models are applicable (see Fig.~\ref{fig:hrd}).  A key
result in this paper is that the surrounding environments of most of
the LBVs don't match expectations for stars with such high $M_{\rm
  eff}$, indicating that single-star models are probably inapplicable.

After contemplating Figures~\ref{fig:mw}, \ref{fig:lmc}, and
\ref{fig:smc} for a time, one comes away with the qualitative but
nevertheless stunning realization that LBVs are often remarkably
isolated, and that they generally seem to avoid massive clusters where
most O-type stars are concentrated.  A few LBVs are in clusters or
associations and their values of $M_{\rm eff}$ may not seem to be in
obvious conflict with their surroundings, but the majority of known
LBVs are much more isolated than they should be for their value of
$M_{\rm eff}$. (For the $M_{\rm eff}$ values implied for LBVs --- even
the low luminosity ones --- {\it all} LBVs should be closely
associated with mid to early O-type stars on the main sequence.)
Individual objects in these two categories are discussed below.

\subsubsection{LBVs in or near Clusters and OB Associations}

{\it $\eta$ Carinae:} The famous LBV $\eta$ Car is indeed in a young
massive star cluster (Tr16 in the Carina nebula) containing a number
of other massive O type stars (see \citealt{smith06}), so in this
sense its environment seems to make sense for the traditional view of
LBVs --- {\it but it is the only one!}  $\eta$ Car is also the most
extreme case; the initial mass implied by its luminosity is of order
200-250 $M_{\odot}$ (see Table~1), so even if it were the result of a
merger that doubled its mass, both component stars must have been
extremely massive.  At such high initial mass and luminosity, H-core
burning lifetimes of stars converge to be almost the same; below 100
$M_{\odot}$, differences in lifetimes for different main-sequence
masses will be more significant.

{\it HD 5980:} This is the most luminous star in the SMC, which was
considered a WR star until it was observed to suffer an LBV eruption
in the 1990s (see \citealt{km08} for a review).  It is in a close
binary with another WR star, plus a wider companion O star.  It is
usually ascribed as a member of the massive young cluster NGC~346,
which contains a number of early O-type stars.  Upon close
examination, however, HD~5980 is actually located well outside the
cluster, at least 20-30 pc from the cluster center in projection.  It
is about 0.2 deg or at least $\sim$20 pc from any other O-type star
(excluding its own bound companions).

{\it R127:} This is one of the classic high-luminosity LBVs, closely
resembling AG~Car in many respects, and it is the brightest star in
the LMC when at its maximum of the S~Dor cycle.  Aside from $\eta$
Car, R127 is the only LBV that clearly resides in a cluster, but in
this case it is a very small cluster by comparison.  \citet{hm03} have
studied the host cluster of R127 in detail (see also
\citealt{walborn91}); it is a small Trapezium-like group of which only
14 members have been detected, and which shows a large age spread with
some stars that are apparently 6-8 Myr old.  Since R127 has a very
high luminosity and a very high inferred initial mass of $\sim$90
$M_{\odot}$, the divide between R127's presumed mass and that of the
older stars in this very small cluster ($\sim$25 $M_{\odot}$) would be
very surprising in a standard single-star evolutionary scenario.

{\it P Cygni:} This classic LBV is not in a cluster, but it is in the
extended Cyg OB1 association.  This region has a substantial space and
age spread, making an environmental estimate of P Cygni's age and mass
very uncertain (the implied $M_{\rm ZAMS}$ is roughly 15--50
$M_{\odot}$ based on the types of nearby main-sequence stars).
According to its luminosity compared to single-star evolutionary
tracks, P Cyg should have an initial mass of roughly 55 $M_{\odot}$.

{\it S Dor and R85:} These two LBVs are in the OB association LH41
(see \citealt{massey00}), which is broadly similar to the loose
association in which P Cygni resides.  The LH41 association has a very
large age spread, containining early O-type stars as well as evolved
stars with likely initial masses of 10-15 $M_{\odot}$
(\citealt{massey00}; see their Table 6 and Figure 7).  The initial
mass and age of any LBV in this association therefore has a large
uncertainty, potentially extending down to quite low masses.  Based on
their luminosities, S Dor and R85 should have initial masses of 55 and
$\sim$28 $M_{\odot}$, respectively.

{\it HD 168607 and 168625:} These two stars (one LBV and one LBV
candidate) form a suspiciously close pair, found only $\sim$1{\arcmin}
($\sim$0.5 pc) apart on the sky.  HD~168607 is a confirmed LBV, and
HD~168625 is an LBV candidate based on its unusual triple-ring nebula
\citep{smith07} resembling the one around SN~1987A. They are found in
the vicinity of the M17 nebula, but not in its central star cluster
NGC~6618.  They are located $\sim$20-30 pc outside NGC6618, well
separated from the cluster and its associated bright nebulosity.

{\it Wra 751:} Although Wra 751 is projected on the sky amid the loose
Car OB2 association, the implications of its apparent membership are
unclear and its distance is very uncertain.  This region looks down a
tangent in the Carina spiral arm, and several clusters and groups of
OB stars that are unrelated in 3D space are projected near one another
(indeed, Wra 751 is thought to be more distant than most of its
neighboring O stars in Figure~\ref{fig:mw}).  Wra 751 is in a crowded
field.  \citet{pasquali06} discussed a small cluster that they
attributed as the likely birth cluster of Wra~751 based on similar
reddening.  Like some other LBVs discussed here, however, Wra~751 is
separated from the rest of the grouping of massive stars in the
cluster.  It resides about 3{\arcmin} (or at least 5 pc projected
separation at 6 kpc distant) southwest of the small cluster's center,
whereas the other stars claimed to be members have a tight grouping
within $\sim$1{\arcmin} \citep{pasquali06}.  Moreover, while Wra~751's
luminosity implies an initial mass of 50 $M_{\odot}$ or more from
single-star tracks (Table~1), the cluster does not contain any early
or mid O-type stars.  The five brightest and bluest stars in the
cluster identified by \citet{pasquali06} include 3 late O-type stars
(O8 V, O9 V, and O9 I) and two early BSGs with luminosities suggestive
of initial masses closer to 20-25 $M_{\odot}$ and an age older than 4
Myr.  Even if this cluster is the birthsite of Wra~751, it implies a
cluster turnoff mass that is not commensurate with Wra~751's very high
luminosity and mass.

{\it R143:} This luminous LBV has an implied initial mass of $\sim$60
$M_{\odot}$ (Table~1).  It is found in the LH100 association, which is
distinct from and well outside the central R136 cluster in 30 Dor
(about 40 pc away).

%%%%%%%%%%%%%%%%%%%

In summary, even among the LBVs that are projected on the sky near
other O-type stars, only the extreme case of $\eta$~Car is clearly in
a young massive cluster.  R127 is in a tiny cluster with only a couple
other massive stars, some of which are much older than expected for a
coeval cluster if R127 really has $M_{\rm ZAMS}\simeq$90 $M_{\odot}$.
A few LBVs are near but outside a cluster (HD~5980, HD~168607,
HD~168625, Wra~751), and a few others are in loose OB associations
with large age spreads that permit a wide range of initial masses
(P~Cyg, S~Dor, R85, R143).  Even granting the benefit of the doubt
that they are actually associated with nearby O stars projected near
them on the sky, it remains astonishing that these constitute less
than half the known LBVs.

\subsubsection{Relatively Isolated LBVs}

\begin{figure*}\begin{center}
\includegraphics[width=4.6in]{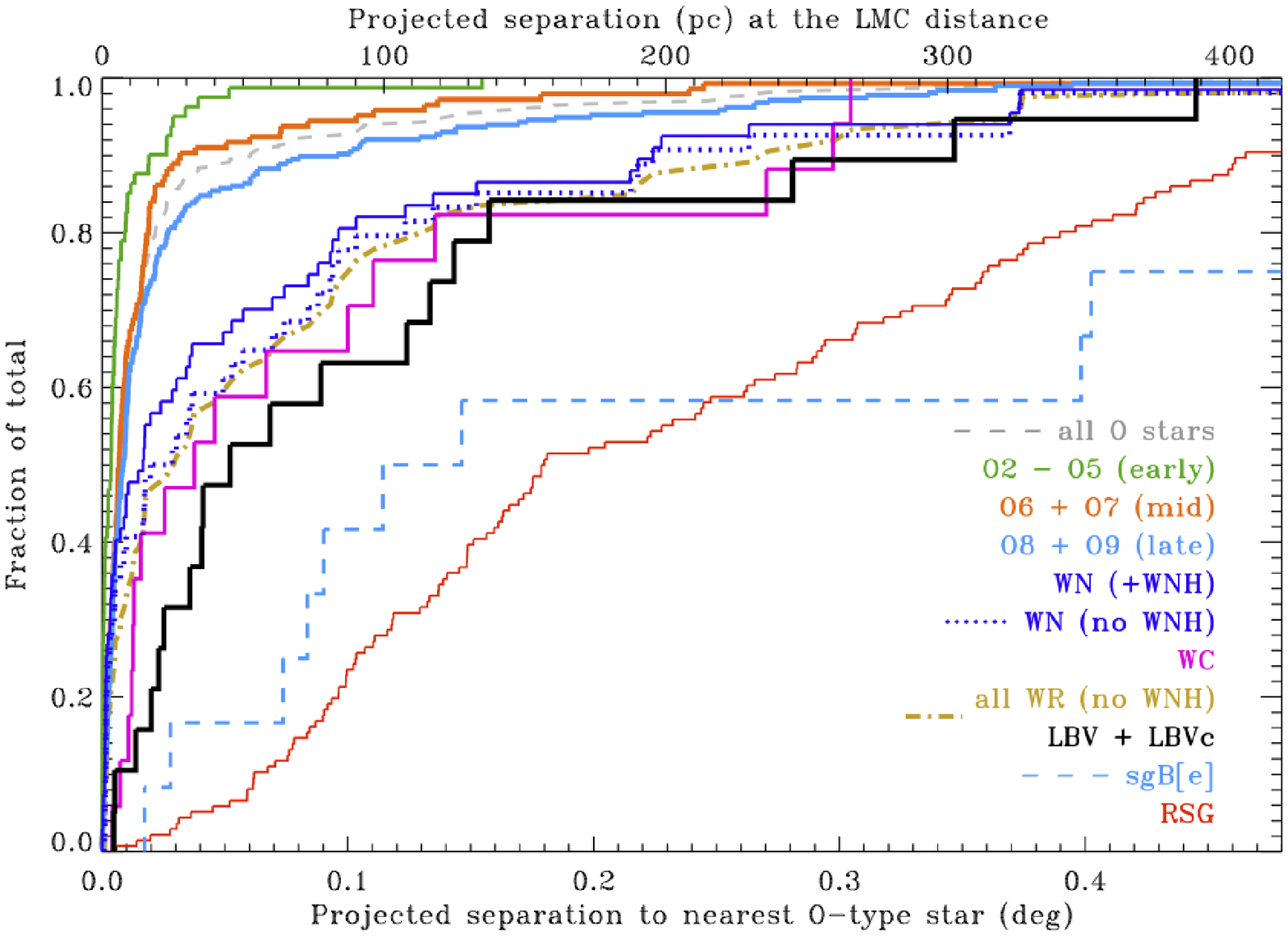}
\includegraphics[width=4.6in]{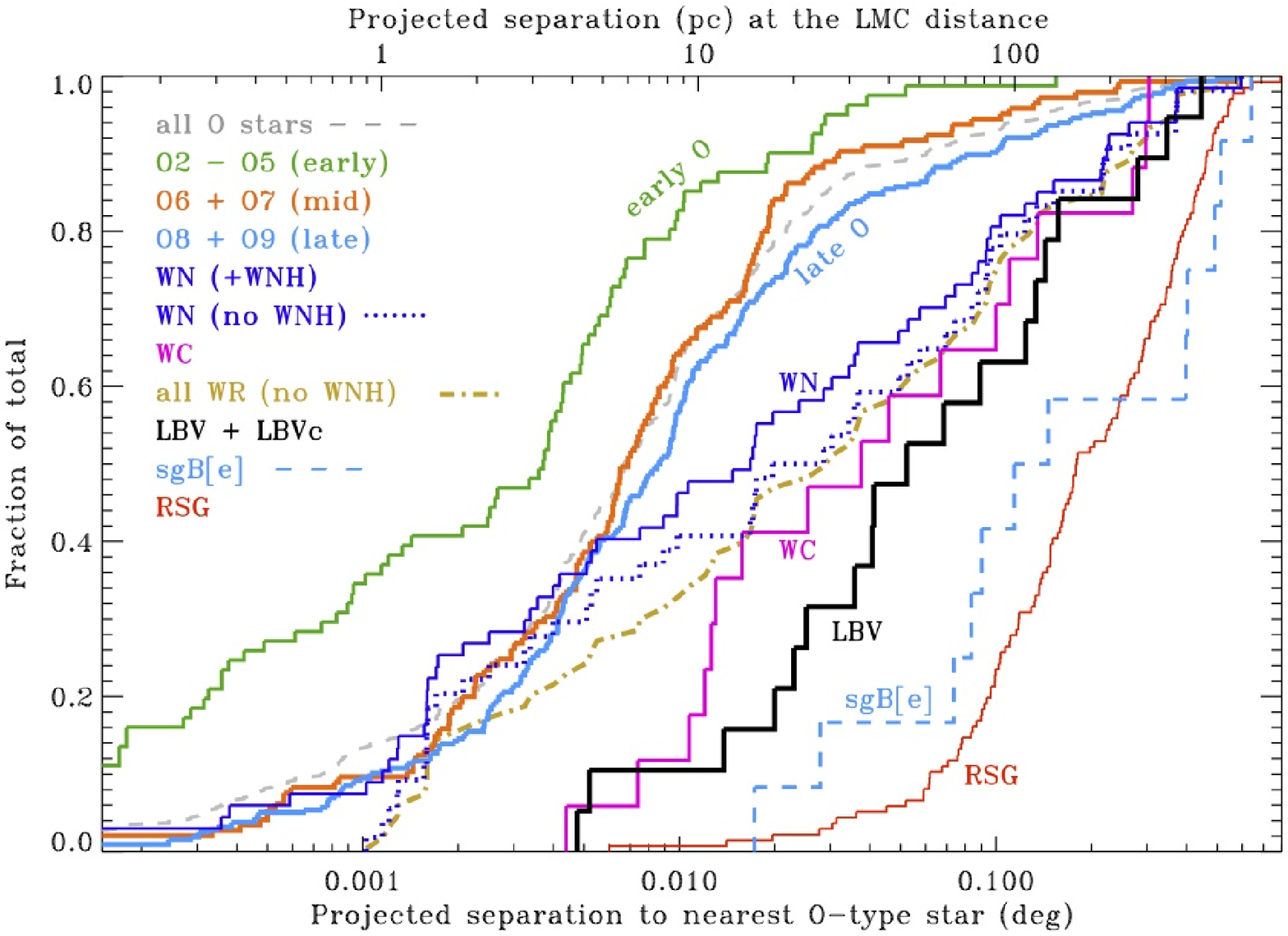}
\end{center}
\caption{Cumulative distribution plot illustrating the differing
  degrees of isolation among various classes of massive stars in the
  LMC; the top and bottom plots are the same, except that the bottom
  has the separation plotted on a log scale. Classes of objects that
  are more clustered with young O-type stars appear farther to the
  left.  The relative ``isolation'' is represented here by the
  distribution of distances to the nearest O-type star (a plot with
  projected distances to the second-nearest O-type star looks very
  similar; Fig.~\ref{fig:cum2}).  For each star in each category, we
  calculated the projected separation on the sky in degrees between
  that individual star and any O-type star (any spectral or luminosity
  class).  Drawn from SIMBAD (except for the LBVs and sgB[e]s), the
  sample includes all O-type stars, WR stars, LBVs, sgB[e] stars, and
  RSGs within a 10$\arcdeg$ projected radius of 30 Dor (except for the
  SMC stars; see text).  The O stars are further subdivided into early
  (O5 and earlier; green), mid (O6+O7, orange), and late (O8+O9, cyan)
  subtypes (these correspond to the same colors of plotting symbols in
  Figures~\ref{fig:mw}, \ref{fig:lmc}, and \ref{fig:smc}.  For WR
  stars, we show WC stars (magenta), a collection of all WN stars
  including WNH stars (solid blue), as well as WN stars without WNH
  (dashed blue).  The mustard dot-dashed line is for all H-poor WR
  stars (WN+WC).  For LBVs, we include both LMC and SMC targets (the
  sparation of the three SMC targets has been multiplied by 1.2 to
  adjust for the difference in distance), and we include both
  confirmed and candidate LBVs (see Table 1).  RSGs (red) are stars
  with spectral types later than K3 and luminosity classes of I, Ia,
  or Iab, and the supergiant B[e] sample (cyan dashed) is from the
  literature \citep{bonanos09,zickgraf06}.}
\label{fig:cum}
\end{figure*}

\begin{figure*}\begin{center}
\includegraphics[width=4.6in]{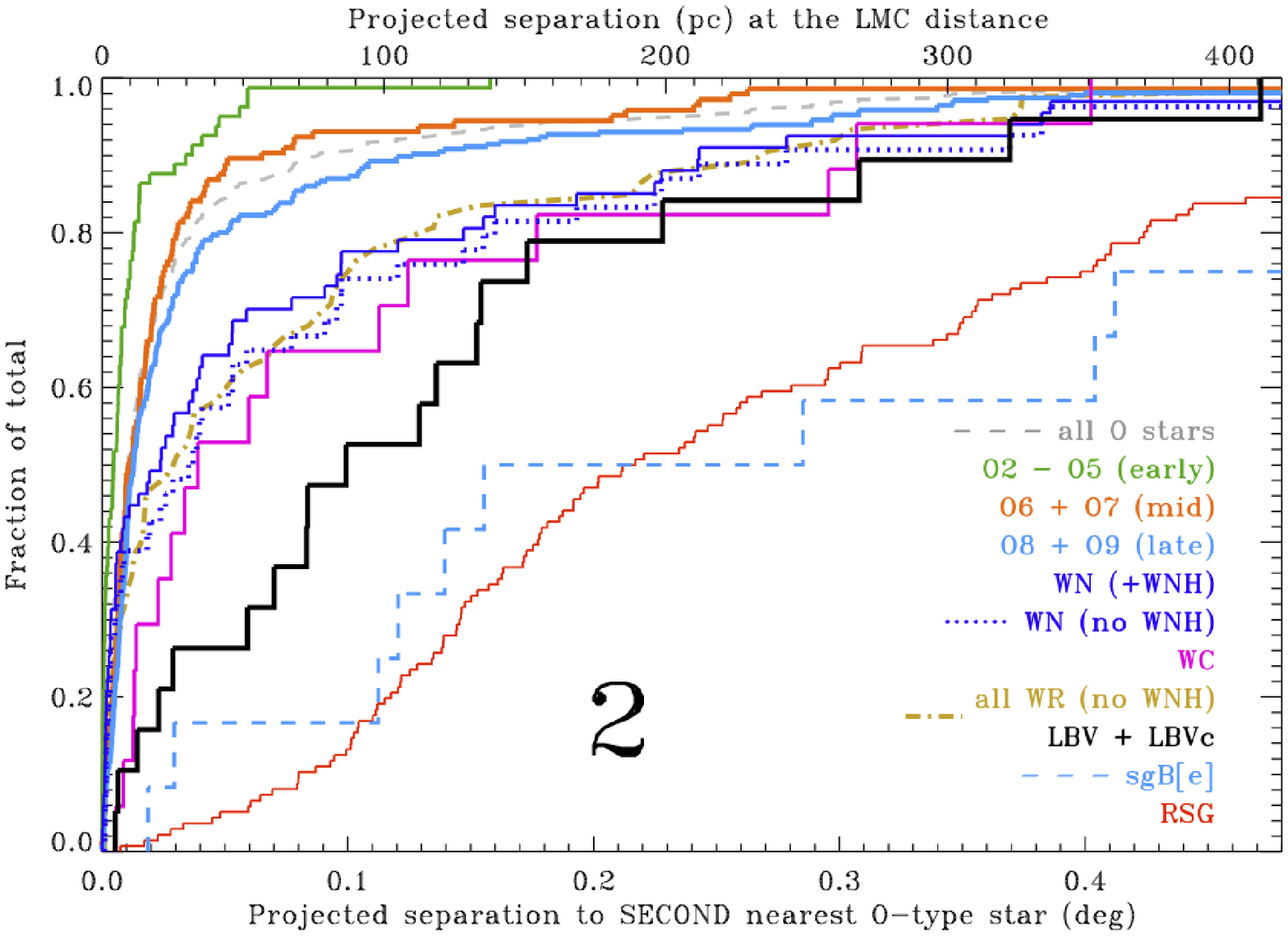}
\includegraphics[width=4.6in]{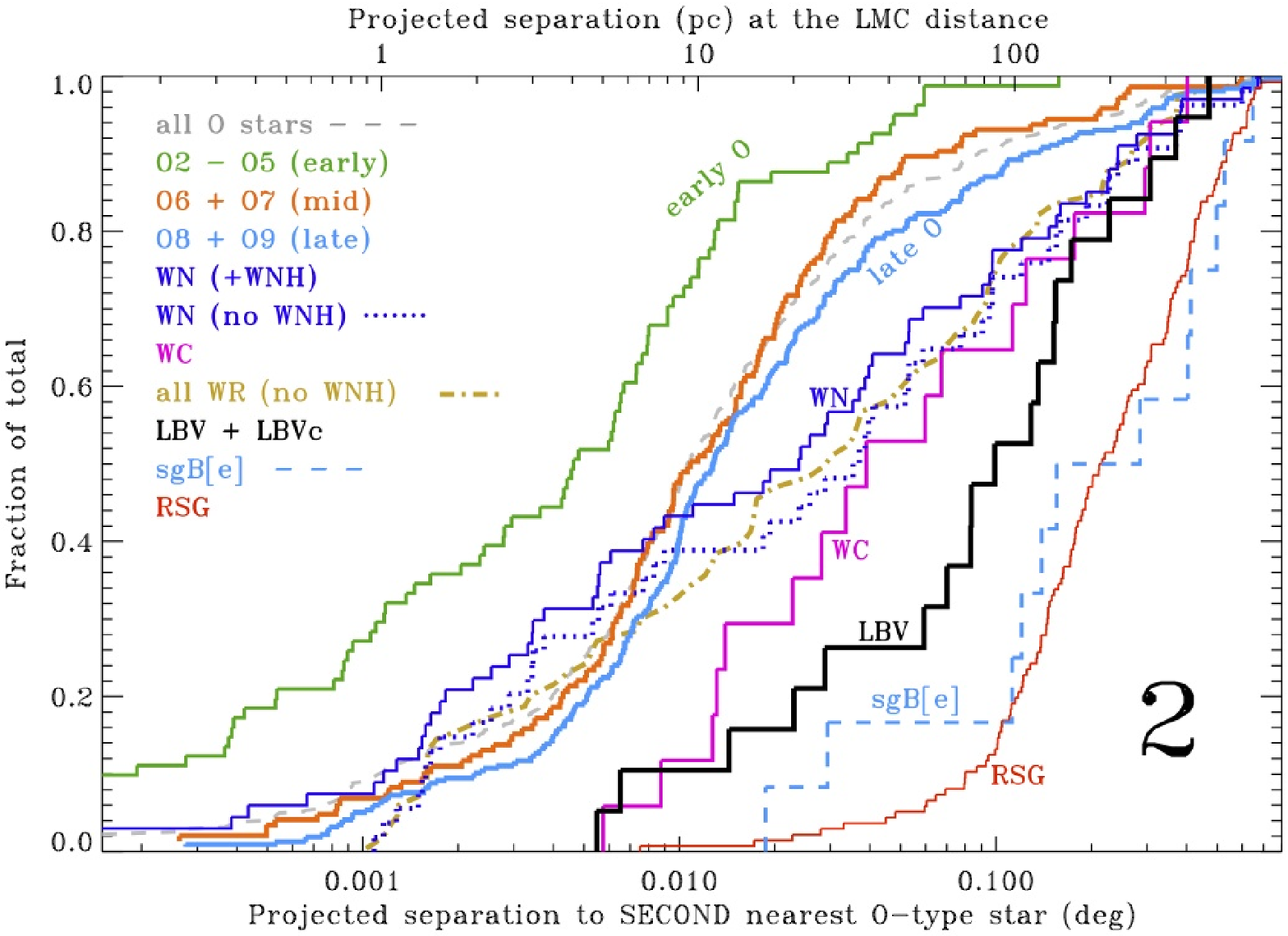}
\end{center}
\caption{Same as Figure~\ref{fig:cum}, but for the projected
  separation to the {\it second}-nearest O-type star (hence the
  ``2'').}
\label{fig:cum2}
\end{figure*}

More than half the LBVs in our sample are in no cluster or OB
assocation at all, and many are 100s of pc from any O-type star.  This
is in contrast to the vast majority ($\sim$80-90\%; see
Figs.~\ref{fig:cum} and \ref{fig:cum2}) of O stars that do reside in
clusters.  Isolated LBVs are discussed here.

{\it AG~Car and Hen~3-519:} Both these very luminous stars are found
in a relative void between the Carina Nebula and the large and diverse
Car OB2 association.  Projected on the sky, AG~Car is more than 0.4
deg from the nearest O-type star, and Hen 3-519 is even farther away.
At a distance of $\sim$3 kpc (corresponding to many of the O stars in
that direction), this would indicate a separation from the nearest O
stars of at least 20 pc, which is surprising considering that AG~Car's
very high luminosity implies an initial mass around 100 $M_{\odot}$.
The distance to AG~Car is highly uncertain, but usually assumed to be
even larger at $\sim$6 kpc \citep{rmh89,groh09,stahl01} based on its
radial velocity and reddening.  This would make it $\sim$45 pc away
from any O star as projected on the sky, and at a larger heliocentric
distance than many of the surrounding O stars.  This distance,
however, may be wrong if AG~Car has substantial peculiar motion.  In
any case, it is much more isolated than expected for a 100 $M_{\odot}$
star.

{\it HR Car:} This LBV has only one O-type star projected on the sky
within 10 pc, and only a few more within 100 pc, all of which are
late-type O stars.  The nearest massive clusters (projected on the
sky) are in the Carina Nebula and Wd1, both of which are well over 100
pc away and at different distances.  With a presumed initial mass of
$\sim$35 $M_{\odot}$, HR Car is therefore surprisingly isolated.

{\it HD 160529 and MWC 930:} These LBVs are also surprisingly isolated
for their luminosity, being found 50-100 pc from any other O star, and
not near any massive cluster.  The nearest O stars are all late-type O
stars that are also relatively isolated.

{\it R110, R81, R71, \& MWC112:} These LBVs are all in the category of
relatively low-luminosity LBVs, sometimes assumed to be post-RSGs (see
\citealt{svdk04}).  Their luminosities suggest initial masses of 25-40
$M_{\odot}$ when inferred from single-star evolution tracks (Table~1).
None of these is in a cluster or association, and in fact each is
$\sim$100 pc or more from any known O-type star.  R81 is about 100 pc
from a small cluster of O-type stars, and the others have one or two
isolated late-O stars as the nearest one.  R110 and MWC112 are
interestingly near to one another on the sky.  R71 is particularly
isolated, with not a single O-type star in the entire plotted window
($\sim$300 pc away) in Fig.~\ref{fig:lmc}.

{\it R40:} Located in the SMC, R40 is not a member of any known OB
association.  It is very isolated, located $\sim$120 pc from any other
O-type star.  Based on its luminosity, it should have an initial mass
of 30-35 $M_{\odot}$.  The lack of any O stars within more than 100 pc
is therefore quite surprising.

In summary, we find that over half the known LBVs are isolated from
other massive stars.  Of the group discussed above, some of the most
luminous classic LBVs are 10s of pc from any other O-type stars, and
many of the ``low-luminosity'' (still quite massive stars with initial
masses of 30-40 $M_{\odot}$) LBVs are 100 pc or more from any other
O-type star.  This degree of isolation is extreme.

While there can be a selection bias against finding fainter O-type
stars near very bright LBVs if they are very distant and reddened
(this is why we have excluded known LBVs with very high reddening such
as those in the Galactic center), there is no conceivable bias that
would prevent one from recognizing bright LBVs in known star clusters.
While there are a couple LBVs in clusters, the avoidance of young
clusters by a majority of LBVs is a robust result.  It seems highly
unlikely that any unobscured massive young clusters surrounding these
LBVs have escaped detection, since they have been studied with deep
high contrast imaging to search for nebulosity, as noted earlier.
Moreover, we do indeed see late O type stars around some LBVs, but it
is the early-type O stars (i.e. the more easily detected ones) that
are missing.  The lack of brighter early O-type stars cannot be a
detection bias when fainter late O-type stars are clearly detected.  A
general conclusion, then, is that LBVs are either extremely isolated
from other O stars, or in cases where there are O stars nearby, those
O stars are consistent with a substantially lower main-sequence
turn-off mass than what we expect for the LBV ($M_{\rm eff}$ in Table
1).  A key qualitative conclusion is that LBVs must either live longer
than allowed by single-star models appropriate for their luminosity,
or they must have selectively traveled far from their birthsites (or
both).  The next section shows that this conclusion holds
quantitatively and is even more clear when we examine a sample of LBVs
that are all at the same distance from us in the Magellanic Clouds.

%%%%%%%%%%%%%%%%%%%%%%%%%%%%%%%%%%%%%%%%%%%%%%%%%%%%%%%%%%%%%%%%%%%%%%%
%%%%%%%%%%%%%%%%%%%%%%%%%%%%%%%%%%%%%%%%%%%%%%%%%%%%%%%%%%%%%%%%%%%%%%%

%%%%%%%%%%%%%%%%%%%%%%%%%%%%%%%%%%%%%%%%%%%%%%%%%%%%%%%%%%%%%%%%%%%%%%%
%%%%%%%%%%%%%%%%%%%%%%%%%%%%%%%%%%%%%%%%%%%%%%%%%%%%%%%%%%%%%%%%%%%%%%%

\subsection{Statistical Distributions}

In this section we quantify the association (or lack thereof) between
LBVs and other massive stars by examining the cumulative distribution
of their projected separations on the sky.  For this statistical
analysis, we focus on the LBVs in the LMC (and the SMC, although there
are only 2 LBVs and 1 LBV candidate there).  We cannot use the
population of LBVs in the MW for this analysis, because the MW is
plagued by uncertainties in distance along the line of sight, as well
as uncertain extinction in sightlines through the disk.  This is
problematic where O stars and LBVs along the same line of sight might
not actually be related in 3D space, and where reddening and
extinction may introduce serious selection effects that would strongly
bias a statistical analysis.  These effects are minimized in the LMC
and SMC, where all the stars are at roughly the same distance, and
where we are not looking along the plane of a disk.

\subsubsection{Constructing the Cumulative Distribution Plots}

The results of our analysis are shown in Figs.~\ref{fig:cum} and
\ref{fig:cum2}, which present the cumulative distribution of projected
separations on the sky between various types of massive stars and
O-type stars.  First, we compiled a list of all O-type stars within
10$\arcdeg$ of 30 Dor as listed in SIMBAD.  We then calculated the
projected separation on the sky between every O star and its nearest
neighboring O star of any subtype, as well as the second-nearest.
Fig.~\ref{fig:cum} shows cumulative distributions of this separation
for the nearest O star; Fig.~\ref{fig:cum2} is the same but for the
second nearest.  (The significance of adding the second-nearest O-type
star is discussed below.)  The O stars are broken into 3 bins for: (1)
early O-type stars (green; O5 and earlier with any luminosity class),
(2) for mid O-type stars (orange; O6-O7 with any luminosity class),
and (3) late O-type stars (cyan; O8-O9 of any luminosity class).

Figs.~\ref{fig:cum} and \ref{fig:cum2} reproduce the entirely expected
result that the earliest O star spectral types, corresponding to the
initially most massive stars, are more highly concentrated than all
other types of massive stars ({\it this group is also the most
  significant for this study, because it is the early O-type stars
  that are usually presumed to be the most likely progenitors of
  LBVs}).  Mid and late O-type stars have their distributions skewed
progressively to the right, mainly because these stars have longer
lifetimes than early O-type stars.  The most massive O-type stars die
first, while clusters and associations spatially disperse with time
due to random motions of stars.  Therefore, on average, for
progressively later O-type stars one must travel farther before
encountering another O-type star, matching the distributions in
Figs.~\ref{fig:cum} and \ref{fig:cum2}.

Fig.~\ref{fig:cum} provides a handy indication that massive stars are
born preferentially in clusters, and that the relative concentration
increases with mass: almost 90\% of early O-type stars are found
within 10 pc of another O star, and for later O-types, about 2/3 are
within 10 pc of another O star.  This average separation increases as
stars age.

Next, we did the same analysis for WR stars, supergiant B[e] stars (or
sgB[e]), RSGs, and LBVs.  For each, we calculated the separation
between a given star and the nearest O-type star, as well as the
second nearest.  LMC WR stars within 10{\arcdeg} of the position of
the R136 cluster were retrieved from SIMBAD, and were separated into
WC stars (magenta), and WN stars (both with and without H indicated in
their classification; dashed blue and solid blue, respectively). We
also show all H-free WR stars (mustard dot-dashed line) in the
figures.  The sample of sgB[e] stars includes only 12 objects in the
LMC \citep{bonanos09,zickgraf06}.  For RSGs (red), we included all
stars of spectral types later than K3 and luminosity classes of I, Ia,
or Iab as listed in SIMBAD for the same 10$\arcdeg$ radius around 30
Dor.  The LBVs correspond to the LBVs and LBV candidates in the LMC
and SMC from Table~1. (For the LBVs in the SMC, we obviously compiled
a separate list of O stars in the SMC, and adjusted the projected
separation by $\sim$20\% to correct for the SMC's larger distance.)

\subsubsection{WR Stars}

For WR stars, Figs.~\ref{fig:cum} and \ref{fig:cum2} reveal the
expected general result that they are more dispersed than O-type
stars.  Naively, this should be the case because the chemical
compositions of WR stars require them to be post-main-sequence
objects.  The mustard-colored dot-dash line shows all WR stars
excluding WNH stars, which appear shifted to the right compared to all
O-type stars (grey dashed).

Next, we consider WR subtypes.  Examining WN stars with and without H
in their spectra, we find that WN and WNH stars have quite similar
distibutions, except that a tail of very tightly clustered objects
less than 1 pc from an O star are all WNH (even more clustered than
later O-type stars); this is also not surprising if WNH stars are
actually the late core-H burning phases of the most massive O-type
stars (see, e.g., \citealt{smithconti08} or \citealt{crowther07} and
references therein).  WC stars are more spatially dispersed than WN
stars.  While about half of WN stars are within 10 pc of an O-type
star, less than 10\% of WC stars are within 10 pc of an O-type star.

The difference between WN and WC is actually quite interesting.  Based
only on their locations that are more dispersed from O-star clusters
than WN stars, one would conclude that WC stars either originate from
preferentially {\it lower} initial masses than the progenitors of WN
stars, or that WC stars are substantially older and correspond to a
more advanced stage in the same evolutionary path that occurs after
most WN stars.  This relative distribution is inconsistent with WC
stars originating from selectively higher initial masses than most
WNs, since H + He burning lifetimes of the most massive stars are not
long enough to migrate 90\% of the population so far away from
neighboring O-type stars.  Stellar evolution models with strong
stellar-wind mass loss that are able to produce WR stars from
single-star evolution make some predictions that are inconsistent with
these results, even ignoring LBVs (discussed next).  With or without
including rapid rotation, single-star models typically expect WC stars
to arise from the most massive stars (e.g.,
\citealt{georgy12,heger03}).  If H-free WR stars descend from the most
massive main-sequence stars, then models predict that the time from
birth to the latter half of the WR phase is only about 3-4 Myr.  This
corresponds to the H burning main sequence lifetimes of mid O-type
stars.  If WC stars result from the strongest winds from the initially
most massive stars, then WC stars should show a typical separation
from other O-type stars that is comparable to middle or late O-type
stars of the same absolute age.  Unless there are severe selection
effects that prevent the identification of WC stars in clusters, the
isolation of WC stars in Figs.~\ref{fig:cum} and \ref{fig:cum2} would
seem to falsify one of the central ideas of single-star evolution
models.  This deserves more attention and modeling of cluster
dispersal as a function of delay time distributions, which is beyond
the scope here.  As we will see next, LBVs make things even worse for
single-star evolution models.

%%%%%%%%%

\subsubsection{LBVs}

A very unexpected result --- and perhaps the most significant
empirical finding in this paper --- is that LBVs are {\it even more}
isolated than WR stars.  The distribution of separations from LBVs to
the nearest O star is skewed significantly to the right as compared to
WN stars, and comparable to (but apparently even somewhat more
isolated) than WC stars.\footnote{Note that the distributions for LBVs
  and all types of WR stars converge at the largest separations around
  0.3 deg or $\sim$300 pc.  This merely reflects the fact that within
  a few degrees around 30 Dor, the surface density of late O-type
  stars (see Fig.~\ref{fig:lmc}) is such that one cannot travel very
  far without having a chance alignment with an O-type star.}  This is
important, because it cannot be reconciled with the standard paradigm
of massive single-star evolution (see below), wherein LBVs are an
intermediate evolutionary phase between massive O stars and WR stars.
The distribution of separations between LBVs and the nearest
neighboring O-type stars is shown by the black solid histogram in
Fig.~\ref{fig:cum}, and to the second-nearest O star in
Fig.~\ref{fig:cum2}.  Both plots yield similar results.  The
implications of this relative isolation of LBVs are discussed more
below.

We include two different plots showing the distributions of
separations to the nearest O star and to the second-nearest O star,
because this provides independent information, and helps mitigate the
influence of incompleteness in the O star sample.  It also provides
additional information about how clustered the nearest O stars are,
and hence, quantitative indication that LBVs avoid clusters.  In
general, if massive stars are all in clusters, then Fig.~\ref{fig:cum}
and \ref{fig:cum2} should be basically the same but with everything
skewed by a small amount to the right in Fig.~\ref{fig:cum2}.  This is
true for most types of stars, but looking closely, one can see that
LBVs actually ``pull away'' from O stars and WR stars when looking at
the second-nearest O star.  This tells us something important about
the O stars that are the nearest neighbors of LBVs --- i.e. that {\it
  they are relatively isolated too}.  Most O stars are in clusters,
and so the second-nearest neighbors of most massive stars are also in
clusters.  However, the fact that LBVs pull away in
Fig.~\ref{fig:cum2} means that the nearest neighboring O-type star to
an LBV is typically not in a cluster, but is itself a relatively
isolated O star (this may indicate that a number of these are chance
projections).

Using the distributions of separations from O stars for the various
classes shown in Figs.~\ref{fig:cum} and \ref{fig:cum2}, we
performed Kolmogorov-Smirnov (KS) tests to quantify the statistical
significance of these separations.  Table~2 shows the results, with
the KS test P-value for the distributions of separations to the
nearest (P1) and second-nearest (P2) O-type star.  Each line in the
table compares LBVs to that class of star.  We see that there is
basically no chance that LBVs and any O-type stars are drawn from the
same parent population. The P-values for WR stars are quite
interesting.  For P2, we find a strong indication that LBVs are not
drawn from the same population as all WR stars together, WNH stars, or
WN stars (all these are less than 5\%), and we find a reasonable
probability that LBVs and WC stars have the same parent distribution
of separations.  For P1, LBVs and WNH stars are clearly different, all
WR stars and WN stars show weaker probably of being from different
parent distributions, whereas WC stars are again consistent with LBVs.
The lifetimes and temporal sequences in standard single-star
evolutionary models don't match these observed distributions of the
majority of massive stars.  In those models, LBVs are transitional
objects between O stars and WN stars, and so LBVs should have a
spatial distribution on the sky that is intermediate between these
two.  Instead, LBVs seem most similar to WC stars, which must occur
much later in the single-star evolutionary sequence, and which have
very different chemical composition.  Thus, the idea that the
initially most massive O-type stars evolve to become LBVs and then
pass on to become WN stars followed by WC stars is ruled out with high
confidence, based on where they reside.

Selection effects are a concern in any cumulative distribution plot,
so one can ask what would be needed to make LBVs appear where they
``should be'' (between O stars and WN stars) in Figs.~\ref{fig:cum}
and \ref{fig:cum2}.  For this, we would need an additional number of
LBVs equal to all those LBVs and LBV candidates currently known, but
all these extra LBVs would need to reside in young massive star
clusters and to have escaped detection.  For any turnoff mass, LBVs
will be much brighter at visual wavelengths than O-type stars of the
same bolometric luminosity because they are cooler. As such, LBVs
would be the brightest stars in any such cluster, so having a large
number escape detection seems impossible.

Some LBVs have Ofpe/WN9 spectral types when they are in their
quiescent hot phase, but we see that LBVs have a statistically
different distribution of separations from O stars as compared to WN
stars with H.  This seems to indicate that the population of WN stars
with H in their spectra is a mixed bag that may include stars in very
different evolutionary stages. Clarifying this is left to future work.

It is quite possible that the census of O-type stars in the MCs is not
100\% complete, but this cannot invalidate our results concerning
LBVs. If a more detailed census were to reveal new O-type stars in the
field of the LMC, one would expect the cumulative distribution of LBV
separations in Figures~\ref{fig:cum} and \ref{fig:cum2} to move to the
left, because some of the LBVs would have a closer O-star neighbor
than before.  However, we have compared the WR stars to {\it the same
  population of O-type stars}, so they would move to the left too (and
the distribution of O stars themselves would get tighter).  Thus, the
critical result that LBVs are more dispersed on the sky than WR stars
would remain.  The only way an incomplete census of O stars would
change the result would be if undetected O stars are preferentially
located very close to LBVs and not close to other types of evolved
stars, which brings us back to the the point above that LBVs would
need to reside in clusters of O stars that have gone undetected
despite deep high contrast imaging.

\subsubsection{RSGs and sgB{\rm [e]}}

Unsurprisingly, RSGs are the most isolated from O-type stars among the
evolved massive stars plotted here. RSGs arise from stars of initial
masses of $\sim$10-30 $M_{\odot}$, but by number the population is
dominated by the lower end of this range where stars have long
lifetimes of around 20-100 Myr.  The fact that LBVs are clearly to the
left of RSGs on Figs.~\ref{fig:cum} and \ref{fig:cum2} therefore
indicates that most LBV initial masses are (very roughly) above 12-15
$M_{\odot}$.

\begin{figure*}\begin{center}
\includegraphics[width=5.6in]{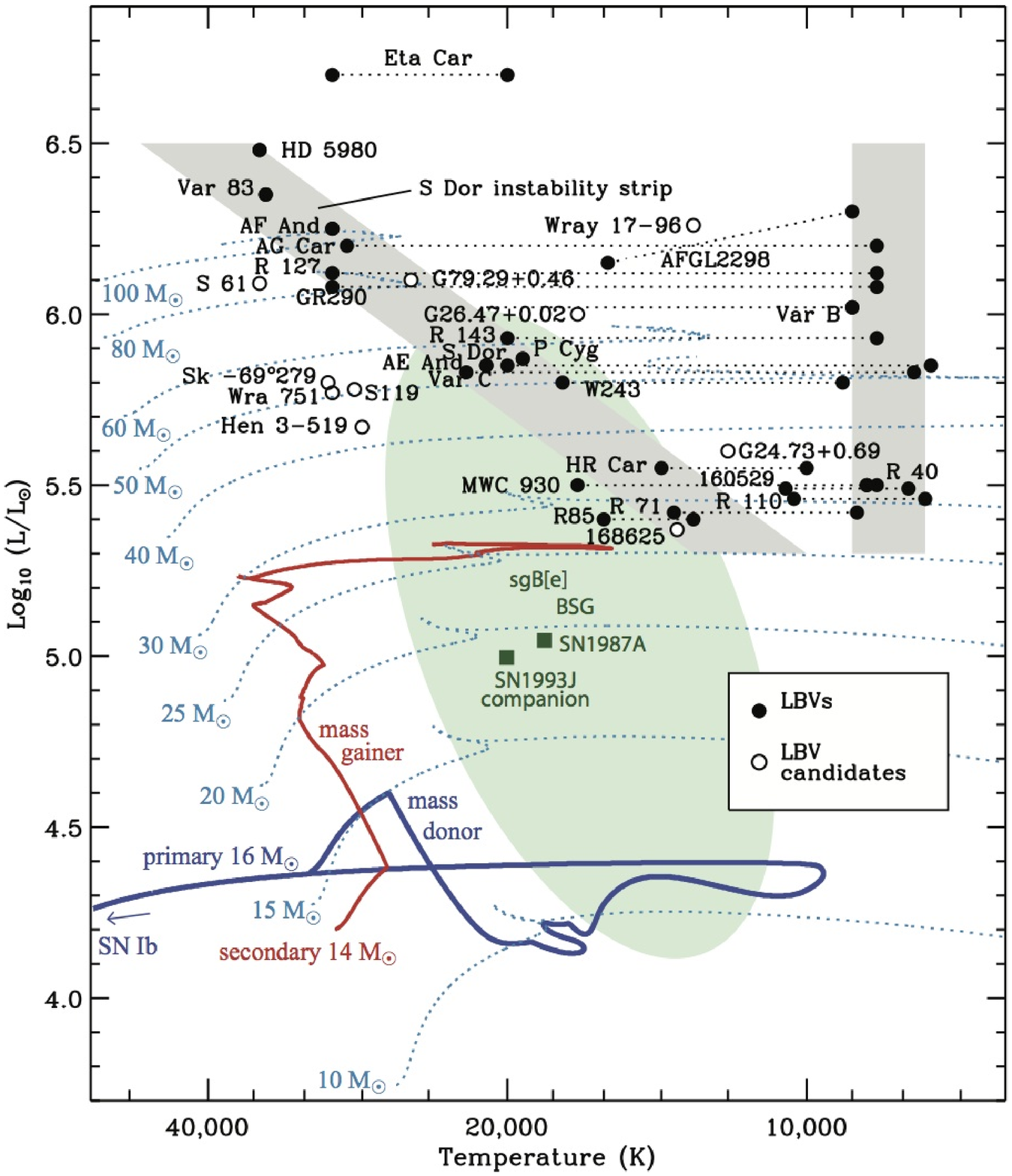}
\end{center}
\caption{HR Diagram comparing LBVs and model evolutionary tracks.  The
  LBVs plotted here are the same as in \citet{svdk04}, except that the
  upper range is expanded to include $\eta$ Car, IRAS 18576+0341 (AFGL
  2298) has been updated \citep{clark09}, and both MWC 930 and R85
  have been added.  \citet{miro14} have claimed that the Galactic star
  MWC~930 is now a confirmed LBV, and \citet{massey00} have argued
  that R85 in the LMC should be counted as an LBV.  The evolutionary
  tracks are from Figure~4a in \citet{lk14}, although as they note,
  the single-star tracks up to 60 $M_{\odot}$ are originally from
  \citet{brott11}.  The blue dotted tracks are single-star evolution
  tracks for $Z_{\odot}$ and an initial rotation speed of 100 km
  s$^{-1}$.  The solid blue and red tracks are for a binary system
  that undergoes RLOF on the main sequence, with an initially 16
  $M_{\odot}$ mass donor and an initially 14 $M_{\odot}$ secondary
  mass gainer.  This illustrates just one example of how a star that
  initially has a relatively low mass can end up as a much more
  luminous star that could resemble the low-luminosity LBVs; binary
  systems with higher initial masses might obviously populate the more
  luminous LBVs in a similar manner. For reference, the approximate
  locations of supergiant B[e] stars and BSGs are shown, as are the
  progenitor of SN~1987A and the putative companion of SN~1993J's
  progenitor \citep{maund04,fox14}.}
\label{fig:hrd}
\end{figure*}

Somewhat more interesting is that the supergiant B[e] stars (sgB[e])
in the LMC (dashed cyan) have a spatial distribution on the sky that
appears very similar to that of RSGs.  We used the positions for the
12 sgB[e] stars known in the LMC \citep{bonanos09,zickgraf06}, and
their separations to the nearest and second nearest O-type stars are
shown along with the other types of stars in Figs.~\ref{fig:cum} and
\ref{fig:cum2}.  A KS test comparing the distributions of sgB[e] stars
to RSGs in the LMC reveals P1 and P2 values of 0.49 and 0.89
respectively (not shown in Table~2), indicating a high probability
that for these two classes of evolved stars, the spatial distributions
relative to O-type stars are drawn from the same population.  A KS
test gives a somewhat weaker indication that sgB[e] stars and LBVs are
drawn from the same distribution (the P1 and P2 values comparing LBVs
to sgB[e] stars are given in Table~2).  The sgB[e] stars look more
isolated than LBVs in Figures~\ref{fig:cum} and \ref{fig:cum2}, but
their low numbers (only 12 in the LMC) make a statistically
significant determination difficult.  Although the luminosities of the
most luminous sgB[e] stars overlap with the lower luminosity LBVs
(see, e.g., Figure~\ref{fig:hrd}), their distribution of luminosities
is skewed lower than LBVs.  Given these considerations, it seems quite
likely that the sgB[e] stars are the evolutionary analogs of LBVs at
somewhat lower initial mass (e.g., where the results of instability
may be less violent).  This is discussed more below.

%%%%%%%%%%%%%%%%%%%%%%%%%%%%%%%%%%%%%%%%%%%%%%%%%%%%%%%%%%%%%%%%%%%%%%%
\begin{center}
\begin{table}\begin{minipage}{2.8in}%
    \caption{KS tests comparing LBV separation distributions to those
      of other star types.}\scriptsize
\begin{tabular}{@{}lll}\hline\hline
Type        &P1   &P2 \\ \hline
%%%
early O  &5.5e-9        &7.4e-8   \\
mid O    &1.4e-7        &1.1e-6   \\
late O   &4.4e-6        &1.2e-5   \\
WR (WN+WC) &0.072       &0.024    \\
WN+WNH   &0.0095        &0.0093   \\
WN       &0.057         &0.036    \\
WC       &0.54          &0.33     \\
sgB[e]   &0.12          &0.22     \\
RSG      &0.00077       &0.0055   \\
%%%%
\hline
\end{tabular}
%%%%%%%%%%%%%%%%%%%%%%%%%%%%%%%%%%%%%%%%%%%%%%%%%%%%%%%%%%%%%%%%%%%%%%%%%
%%%%%%%%%%%%%%%%%%%%%%%%%%%%
\medskip

Notes:  \\
P1 = KS P-value for separations to the nearest O star. \\
P2 = KS P-value for separations to the second-nearest O star. \\

%%%%%%%%%%%%%%%%%%%%%%%%%%%%%%%%%%%%%%%%%%%%%%%%%%%%%%%%%%%%%%%%%%%%%%%%
\end{minipage}\end{table}\label{tab:ks}
\end{center}

\subsection{Summary of LBV Environments}

The large fraction of LBVs that appear isolated should sound alarm
bells for anyone familiar with the traditional view of LBVs as
``massive stars in transition'', just having finished core-H burning
and soon moving on to He burning as WR stars.  Upon comparing
positions of O-type stars to the coordinates of known LBVs, the result
is astonishing. Basically, {\it LBVs systematically avoid clusters of
  O-type stars, and they are rarely associated with O-type stars of
  similar (presumed) initial mass}.  This is the opposite of what is
expected in the standard paradigm of massive star evolution where LBVs
are the immediate descendants of the most massive O stars, and the
predecessors of luminous WR stars.  Figs.~\ref{fig:cum} and
\ref{fig:cum2} indicate that the vast majority of O-type stars are
indeed in highly clustered environments.  WR stars are significantly
more dispersed, consistent with their more evolved status.  LBVs
should be intermediate between O stars and WR stars, but in fact they
are more isolated than WR stars.  This requires that they are older,
on average, than WR stars, which in turn implies that they arise
either from lower initial masses (longer core-H burning lifetimes),
that they are a more advanced evolutionary state than WR stars
(improbable based on chemical composition), or that some mechanism has
systematically removed them from clusters without disturbing other
massive stars.

Initial masses implied by LBV environments differ considerably from
other inferences about their likely initial masses based primarily on
turnoff masses and ages of nearby stars.  (Actually, there have been
very few studies of the ages and masses of LBVs based on their
environments, perhaps reflecting the fact that many of them are not in
clusters where this is usually done.)  $\eta$ Car is located in the
Tr16 cluster. Tr16 is inferred to have an age of $\sim$3 Myr, which
is, however, based largely on the fact that $\eta$ Car itself has not
yet gone SN.  Excluding $\eta$ Car, Tr16 has a turnoff mass of roughly
80-100 $M_{\odot}$, which is significantly below the presumed initial
mass of $\eta$ Car itself --- although perhaps consistent enough that
it does not immediately raise questions, due to the similar
main-sequence lifetimes of very massive stars at $\ga$100 $M_{\odot}$.
Based on studying OB associations in the LMC/SMC and adopting
characteristic turnoff masses for the earliest O-type stars in each,
\citet{massey00} inferred initial masses of LBVs and WR stars.  They
favored an initial mass of $>$85 $M_{\odot}$ for the two LBVs in their
study (S~Dor and R85), and high initial masses for WR stars as well
($>$70 $M_{\odot}$ in the SMC, and 30-60 $M_{\odot}$ in the LMC).
There are, however, two key effects that compromise this method of
determining initial masses: (1) It requires that the cluster or
association is coeval, but many of the regions studied have large age
spreads compared to the lifetimes of very massive stars.  For example,
the association LH41 in which S~Dor and R85 reside has a very large
age spread; while its most luminous members do suggest a turnoff mass
around 85 $M_{\odot}$ by comparison to single-star evolutionary
tracks, this same association also contains cool supergiants with
initial masses of only 10-15 $M_{\odot}$ \citep{massey00}. (2) This
method ignores important effects of binary evolution that corrupt the
inferred turnoff mass (see, e.g., \citealt{larsen11}).  Both binary
mass transfer and binary mergers will produce luminous stars that
repopulate the upper main sequence (e.g.,
\citealt{lk14,langer12,schneider14,demink14}).  When compared to
isochrones of single-star evolutionary models, these massive blue
stragglers will imply an age younger than the true age of the cluster,
and will overestimate the main-sequence turnoff mass.  Convolved with
an age spread due to star formation that may be comparable to teh
lifetimes of massive stars, this makes it difficult to use cluster
turnoffs to estimate the initial masses of the most luminous evolved
stars.

Looking at the spatial relationship between LBVs and other massive
stars gives an independent and quantitative check on relative ages
that does not have the same pitfalls as cluster turnoffs.  If we
assume that the vast majority of massive stars form in clusters
(indeed they do, as O-type stars are observed to be highly clustered
in Figure~\ref{fig:cum}), then as the most massive stars explode early
and the cluster spreads out due to random motions, the distribution of
distances to a nearest neighboring O-type star must slowly and
steadily increase with age.  This method has no reliance on uncertain
input physics of stellar evolution models or their neglect of
binaries.  It does not give a precise absolute age, but it does
provide a reliable measure of the {\it relative} ages of different
populations of massive stars.  This method produces a very different
result from cluster turnoffs \citep{massey00}.

The relative isolation of LBVs is apparent in the MW as well as the
LMC/SMC, although only the LMC/SMC produce a meaningful statistical
analysis due to distance and reddening uncertainties in the MW.  If
there were only a few isolated LBVs, it wouldn't be so troubling
because we might expect a few cases where the LBV was initially the
most massive star in a small cluster that contains no other O-type
stars.  Indeed, a few percent of O stars (mostly late O-type stars)
appear to have been born in relative isolation (e.g.,
Fig.~\ref{fig:cum} and \citealt{oey13}) --- but $\sim$90\% of early
O-type stars {\it are} in clusters.  LBVs are the opposite: only rare
exceptions are found in clusters ($\eta$ Car and R127) and most of
them are either in loose associations with large age spreads, or they
are found 10s or even 100s of pc from other O-type stars.  This
isolation requires that LBVs are either very massive stars that for
some reason were preferentially born well outside clusters (hard to
justify physically), or that they have lived longer than we expect
based on single-star evolution and have systematically moved away from
their birth environments.  This cannot be reconciled with the
traditional view of LBVs in single-star evolution.

%%%%%%%%%%%%%%%%%%%%%%%%%%%%%%%%%%%%%%%%%%%%%%%%%%%%%%%%%%%%%%%%%%%%%%%
%%%%%%%%%%%%%%%%%%%%%%%%%%%%%%%%%%%%%%%%%%%%%%%%%%%%%%%%%%%%%%%%%%%%%%%

\section{STANDARD VIEW: LBVS AS THE TRANSITIONAL DESCENDANTS OF MASSIVE SINGLE O-TYPE STARS}

As noted in the introduction, the current standard paradigm of
massive-star evolution is that single massive O-type stars evolve into
H-deficient WR stars by virtue of their own radiation-driven mass
loss.  For all but the most luminous and most massive stars
(i.e. $\ga$100 $M_{\odot}$) that pass through a strong-winded WNH
phase, accounting for clumping effects indicates that line-driven
stellar winds are too weak to accomplish this (see \citealt{smith14}
and references therein).  Thus, if LBVs are presumed to fit into the
evolutionary sequence of single stars:

\smallskip 

\noindent O star $\rightarrow$ Of/WNH $\rightarrow$ LBV $\rightarrow$
WN $\rightarrow$ WC $\rightarrow$ SN Ibc,

\smallskip

\noindent then enhanced eruptive LBV mass loss would be essential in
order for single massive O-type stars to become WR stars
\citep{smithowocki06}.  In this scenario, the LBV phase is a very
brief, fleeting transitional phase (see, e.g., \citealt{ln02,groh14}).
In evolutionary models, there is little time ($\sim$10$^5$ yr) after
the end of core-H burning and before the onset of the WR phase when a
star can be an LBV.  In this short time interval, stars cannot move
very far, and so LBVs are expected to have a spatial distribution on
the sky that is very similar to the O stars that are supposedly their
immediate progenitors.  For a typical velocity dispersion in a cluster
of a few km s$^{-1}$, a single star will move less than $\sim$10 pc in
3 Myr, and much less than 1 pc in 10$^{5}$ yr.  Therefore, if early
O-type stars are mostly in clusters (they are; see Figs.~\ref{fig:cum}
and \ref{fig:cum2}), then their immediate single-star descendants must
be as well. In Fig.~\ref{fig:cum}, the cumulative distribution of LBV
separations should therefore be in between O-type stars and WN stars,
and LBVs should be much more clustered than WC stars. In this paper we
have shown that the opposite is true -- that LBVs are surprisingly
isolated from O stars.  Most critically, observations indicate that in
terms of their separation from O-type stars, {\it LBVs are even more
  isolated than WR stars (both WN and WC).}

The observed population of LBVs therefore cannot evolve into the
observed population of WR stars, because LBVs would need to turn
around and systematically move back toward clusters and associations
in order to match the locations of WR stars.  This cannot be.  The
standard view of LBVs as single ``massive stars in transition'' is
incompatible with observations of their locations, and must no longer
be considered as a viable evolutionary scenario for the majority of
massive stars.  This rules out the monotonic evolutionary paradigm
outlined above, making it difficult for single massive stars to become
H-poor WR stars and SNe~Ibc (because without LBV mass loss, their
winds aren't strong enough).  Most often, the LBV phase must play a
different role.  A straightforward alternative is discussed next.

\section{ALTERNATIVE: MASS GAINERS IN RLOF}

The fact that LBVs are more isolated than WR stars, plus the fact that
some stars appear to remain in an LBV phase until death as SNe~IIn
(see \citealt{smith14} for a review), suggests an entirely different
picture.  Instead of the monotonic evolutionary scheme for single
stars discussed above, the spatial distributions discussed in this
paper suggest that massive star populations and especially LBVs are
dominated by bifurcated evolutionary trajectories:

\smallskip
\noindent O star $\rightarrow$ $\begin{cases} {\rm WN} \rightarrow
  {\rm WC} \rightarrow {\rm SNIbc} & {\rm (donor)} \\ {\rm LBV/B[e]}
  \rightarrow {\rm SNIIn} & {\rm (gainer)} \end{cases}$.

\smallskip

\noindent In this scenario, O-type stars evolve off the main sequence,
and through binary interaction the majority of massive stars either
(1) lose their H envelope through mass transfer to a companion, become
WR stars and die as stripped-envelope SNe, or (2) gain mass from a
companion, become BSGs, sgB[e]s, and LBVs, and then retain their H
envelopes until they die as SNe~IIn.  After the mass transfer phase,
the subsequent evolution of the mass gainer may be quite varied, and
so not all mass gainers will necessarily be LBVs and SNe
IIn.\footnote{For example, in some studies, the spun-up mass gainers
  are assumed to undergo quasi-homogeneous chemical evolution and die
  as H-free WR stars and possibly gammay ray bursts (GRBs), as
  described by \citep{eldridge11}.  This is discussed more in
  Section~5.  In some cases the evolution can be quite complicated, if
  for example, multiple episodes of mass transfer or late mergers
  occur.  This may be important for some fraction of massive stars.}
In an insightful early paper, \citet{kg85} discussed the fact that a
few of the Hubble-Sandage variables in M31 appeared relatively
isolated, and they suggested a similar binary mass-transfer scenario
for these, while still favoring a single-star evolutionary scheme for
most LBVs.  Here, we show that the environments of {\it most} LBVs
violate the single-star picture, not just a few exceptions.  Instead
of LBVs being a brief transitional phase for all very massive stars,
they become the dominant late evolutionary phase for a subset of
massive binaries.  LBVs are rare enough that the fraction of stars
which do this and for how long they remain in the LBV phase is poorly
constrained observationally.  (Note that when the oversimplified,
monotonic evolutionary scheme is abandoned, the number ratio of LBVs
to O-type stars offers no meaningful constraint on the LBV duration
without additional information.)

Fig.~\ref{fig:hrd} shows the locations of LBVs on the HR diagram,
mostly taken from \citet{svdk04}, except as noted in the caption.  The
locations of LBVs are compared to standard evolutionary tracks for
single stars, and also to an example of an evolutionary track for an
interacting binary system from \citet{lk14}.  Using single-star
evolution tracks as a reference, the initial masses of LBVs would
appear to range from 30-40 $M_{\odot}$ for the lower-L LBVs, and from
50-250 $M_{\odot}$ or more for the classical high-luminosity
LBVs.\footnote{Note that the values of $M_{\rm eff}$ we infer from
  Figure~\ref{fig:hrd} are approximate.  Values of the luminosity
  depend on the analysis technique used for each star and the distance
  assumed, and there are multiple values available in the literature
  for several LBVs (see, e.g., \citealt{groh09}).}  A main point of
our paper is that the isolated environments of LBVs are incompatible
with the short lifetimes experienced by stars with such high initial
mass, and they are incompatible with the key idea that most LBVs
continue their evolution to WR stars.

Instead, ascribing LBVs as the mass gainers in interacting binaries
provides an attractive and plausible alternative. As noted earlier,
the plausibility of this basic idea was mentioned long ago
\citep{jsg89}, but did not become the dominant view.  Through the
process of RLOF, the mass gainer can substantially increase its mass
and luminosity to resemble a star that had an initially much higher
mass. One example is shown in Fig.~\ref{fig:hrd}, where a star with an
initial mass of 14 $M_{\odot}$ accretes mass in a binary and ends up
with a luminosity commensurate with a $\sim$26 $M_{\odot}$ star
\citep{lk14}.  It is easy to see that such an evolutionary path could,
in principle, give rise to the low-luminosity group of LBVs.
Similarly, somewhat more massive stars in binaries (initial masses of
30-40 $M_{\odot}$), could accrete mass to appear as 60-80 $M_{\odot}$
classical LBVs.  One can imagine higher mass analogs of this that
could correspond to the classical LBVs like AG~Car and R127.

A crucial point is that this mass accretion and corresponding increase
in luminosity may occur after a long delay, after both stars in the
binary system have lived through much or all of their H-core burning
main-sequence lifetimes. As initially lower mass stars, their core-H
burning lifetimes may be much longer than the main-sequence lifetime
expected for the more massive star that they become.  As such, a mass
gainer may appear younger than the population of stars around it.  In
cases where the discrepancy between the initial mass and $M_{\rm eff}$
is close to a factor of 2, the lifetime can be doubled for moderately
massive stars (15-30 $M_{\odot}$).  For example, the approximate
lifetimes of single stars of 15 and 30 $M_{\odot}$ initial mass are
about 14 and 6.5 Myr, respectively \citep{whw02}. If LBVs fit this
role, then they are essentially evolved massive blue stragglers, not
massive stars in transition.  LBVs originating this way may help solve
a number of issues that have been problematic for a single-star
scenario:

1. As noted above, the mass gainer will have a longer age than
expected for its current mass and luminosity, helping to rectify the
isolation of LBVs discussed in this paper and the initial mass
discrepancy.

2.  In addition to becoming more massive and more luminous, the mass
gainer will also gain angular momentum during RLOF.  This may cause
LBVs to appear as rapid rotators late in life.  A number of LBVs do
exhibit properties consistent with rapid rotation
\citep{smith02,smith03b,groh09b,groh06}.  Moreover, rotation has been
inferred to be quite important in some ideas about LBV instability
\citep{langer98}.  In a single-star scenario, it is difficult to
understand how they can shed large amounts of mass in steady winds
while retaining their angular momentum.  Single-star models generally
predict negligible rotational speeds for the LBV phase
\citep{groh14,mm03}.  This scenario also implies that a large fraction
of the fast runaway stars that are kicked out of clusters by their
companion's SN will be these rapidly rotating mass gainers (i.e. there
must be a fair number of them to account for the observed distribution
of LBVs).  It is not clear yet if this is the case, but further study
of this may provide an important test.

3.  RLOF or other binary interaction may produce very asymmetric CSM,
as seen around many LBVs with bipolar or elliptical nebulae ($\eta$
Car, AG~Car, HR~Car, HD~168625, etc.).  Using spectropolarimetry,
equatorial distributions of CSM around SNe~IIn have also been inferred
\citep{mauerhan14,hoffman08,leonard00}.  Some of the more extreme
observed asymmetries in the CSM may be difficult to achieve with
single stars.

4.  Removing the restriction that LBVs must transition into H-poor WR
stars before they die, mass gainers might retain some of their H
envelopes until death.  This would reconcile the problem of LBVs
exploding as SNe~IIn.

5.  The mass gained from a more evolved companion may be significantly
enriched in nitrogen.  This by itself does not necessarily argue
against a single-star scenario if efficient mixing can bring N to the
star's surface \citep{lamers01}, but the N-enriched nebulae around
LBVs are consistent with N enrichment in a binary mass-gainer
scenario.

6.  Similarly, the accretion of enriched material from a companion
could help explain why LBV relative H/He abundances are similar to
those of WN stars with H (actually between those of very luminous WNH
stars and other WN stars with H; \citealt{smithconti08,langer94}),
even though their locations on the sky are very different.

7. When their stripped-envelope companion explodes as a SN~Ibc or IIb,
the mass-gainer star is likely to receive a kick.  This provides an
attractive explanation for why LBVs would preferentially and
systematically seem to avoid star clusters, and adds to their longer
lifetimes in explaining their observed isolation.  Even if they are
born in clusters and live in associations for several Myr, they may
get kicked out of the cluster.  Because the initially more massive
star will usually explode first, the kicked mass gainer may have a few
more Myr to live after that SN event.  Traveling at 50-100 km
s$^{-1}$, the star can move about 50-100 pc in 1 Myr.  Such motion for
only 1-2 Myr is therefore sufficient to explain the relative isolation
of most of the LBVs discussed here.  It is important to recognize that
the mass gainer may not exhibit the LBV instability immediately upon
accreting mass, but is likely to do so in its own time when it
finishes core-H burning and evolves to become a supergiant.  (In the
mean time, it will be a much hotter O-type star that is fainter in
visual light; thus, we would not expect bright LBVs to be seen at the
positions of most SNe~Ibc, because it may take them another 10$^6$ yr
to evolve into their own supergiant phase.) The delayed onset of the
LBV phase may lead them to preferentially appear outside clusters.
Depending on how long it takes them to die, this may also lead SN
impostor eruptions and the eventual SNe~IIn to appear isolated.

8.  Even if they are a product of binary evolution, LBVs may appear as
single stars if their companion has already exploded.  In some cases,
of course, the close companion may not have exploded yet.  If they
were born in triple systems, they may still be in binaries.

9.  In the MW, LMC/SMC, and in other nearby galaxies, there is a large
population of massive stars that resemble LBVs but that have not yet
exhibited the eruptive LBV instability.  These are usually called
``LBV candidates'' as noted earlier.  \citet{massey07} find an order
of magnitude more LBV candidates than bona fide LBVs in nearby
galaxies, and massive dusty shells around massive stars in the MW
suggest a similarly large number here \citep{wachter10,gvaramadze10}.
Counting only LBVs, their small number compared to O-type stars has
been used to justify a very brief transitional phase (e.g.,
\citealt{hd94}) before becoming a WR star.  However, if LBV candidates
are included in the count, the implied LBV lifetime becomes much
longer (see \citealt{smith14} and references therein).  In the
bifurcated evolutionary scheme discussed above, the post main-sequence
lifetime of the H-rich mass gainer is much longer than the LBV phase
envisioned in single-star models, allowing LBV candidates to be the
same stars as LBVs, if the LBV eruptive instability is only active for
part of that time or in a subset of circumstances.

In addition to mass accretion through RLOF, stellar mergers or more
exotic systems such as blue Thorne-Zytkow-like objects (TZOs) might
also yield some of the same results that are compatible with observed
properties of LBVs (overluminous or young compared to surrounding
stars, rapid rotators, enriched and asymmetric CSM,
etc).\footnote{TZOs are normally expected to be red
  \citep{tz75,tz77}. However, at the high luminosities appropriate for
  LBVs ($\sim$10$^6$ $L_{\odot}$), RSGs do not exist.  Very luminous
  shell burning stars that would otherwise be luminous RSGs are
  thought to reside in the blue because of mass loss and instability
  in their outer envelopes.  The same might apply to the envelopes of
  TZOs in this high-luminosity range, if they exist. This is still
  speculative, but we mention it because blue TZOs cannot be ruled out
  for LBVs without more detailed study.}  The role of mergers in
producing asymmetric CSM, in particular, has been discussed
extensively (see review by \citealt{pod10} and references therein).
Theoretical predictions for TZOs are highly uncertain, and it remains
difficult to rule out the possibility that some LBVs are merger
products or blue TZOs, especially those that may still be in clusters
or associations.  However, lacking a significant kick from a
companion's SN, stellar mergers do not offer a compelling
explanation for why the larger population of LBVs seem to
systematically avoid clusters.

So, what about the fates of single stars or binaries that merge early
on the main sequence?  If LBVs are exclusively or mostly a binary
phenomenon, then it becomes very unclear if WR stars can arise from
single stars.  Perhaps RSG mass loss becomes critical.  Uncertainties
(in most cases overestimates) in mass loss, plus errors in models that
have arisen from comparing single-star models to binary populations,
prohibit existing evolution models from making unique predictions
(again, see \citealt{smith14}).  With all known stellar mass black
holes in binaries, there are few observational constraints on the end
fates (back hole or neutron star; successful exoplosion or quiet
collapse) for truly single high-mass stars.

A key conclusion from LBV environments as compared to other types of
stars is that while LBVs are indeed very massive stars based on their
current high luminosity, they didn't necessarily begin their lives
that way. Although this paper advocates a different origin for LBVs
than is usually assumed, much of the traditionally discussed phenomena
associated with LBVs may still apply.  They are still stars with
extreme mass loss and instability that have eruptive super-Eddington
winds or explosive mass loss (see, e.g., \citealt{owocki04,smith11a}).
What is very different in the evolutionary scheme advocated above is
the path by which the stars have come to be in their current state.
Hence, their initial masses and ages are very different than
previously assumed.  LBVs have been discussed as a normal transitional
state for the most massive stars that have evolved to be in close
proximity to the classical Eddington limit.  Instead, we argue that in
most cases they are stars that have become more luminous, rapidly
rotating, and unstable as a consequence of mass accretion in a binary
system (or perhaps in some cases something more exotic, such as a
merger or blue TZO). Although the scenario suggested here does not
solve the overarching mystery of the physical trigger of LBV
eruptions, the evolutionary path to arrive at this point may be a
critical part of the puzzle (especially where angular momentum and
stellar structure are concerned).  This remains a challenge for future
theoretical work.

A clear message with broader significance is that cluster turnoffs
analyzed with single-star model isochrones will systematically
underestimate the age and overestimate the turnoff mass of clusters
and associations, because the apparent turnoffs are contaminated by
massive blue stragglers.  This method is especially unreliable for
very massive stars, where age spreads in clusters are comparable to
their absolute age.  Similarly, we must be mindful of SN progenitors
and their ``initial mass'', as discussed next.

\section{CONNECTIONS TO SUPERNOVAE AND THEIR HOST ENVIRONMENTS}

Evidence presented in recent years has begun to shift our central
paradigm of massive star evolution.  On the one hand, wind mass-loss
rates are lower than we used to think, while the close binary fraction
among O-type stars is shown to be quite high (roughly 2/3 are
interacting, as noted in the introduction).  The isolation of LBVs now
seems to require a major shift.  The interaction process of RLOF or
mergers has a profound impact on the types of evolved stars that are
observed, and binarity must therefore also have a strong (or dominant)
impact on the distribution of SN subtypes that mark the end fates of
these massive stars --- perhaps even moreso than metallicity dependent
winds and initial mass, as has generally been assumed.

Observed statistics of SN subtypes and other considerations already
argue that binary RLOF and not stellar winds must dominate the removal
of the H envelope for most stripped-envelope SN progenitors
\citep{smith11b}, and binary population synthesis studies have
suggested this for some time
\citep{pod92,vanrensbergen96,dvb98,dvb07,izzard04,cantiello07,
  demink07,eldridge08}.  There are too many SNe~Ibc and IIb than there
are stars initially massive enough to remove their H envelopes via
their own winds.  Including SNe~IIb, the fraction of core-collapse SNe
that have stripped-envelope progenitors is $\sim$36\%, whereas stars
that are luminous enough to shed their H envelopes through their own
winds comprise only 10-15\% of a normal SN-producing IMF
\citep{smith11b}, or perhaps even less if some massive stars collapse
to black holes without producing bright SNe.  Interestingly, from the
fraction of O stars that will interact and exchange mass,
\citet{sana12} estimate that 1/3 is the approximate fraction of
massive stars that we might expect to lose their H envelope in RLOF.
Comparing this 33\% to the observed value of 36\% of ccSNe that have
stripped-envelope progenitors, there appears to be little wiggle room
to allow massive single stars to die as SNe~Ibc. Similar arguments in
favor of binary progenitors have been made based on the mass and
composition of the ejecta in stripped-envelope SNe
\citep{dessart12,hachinger12}.

A natural question arises.  If binary RLOF strips the H envelope to
make most SNe~IIb, Ib, and Ic, then what happens to the mass-gainer
companions of these stripped-envelope stars?  What do those
mass-gainer stars look like, and which SNe do they yield?  In a
theoretical study, \citet{eldridge11} assumed that the spun-up mass
gainers would experience enhanced rotational mixing and would undergo
quasi-homogeneous chemical evolution, suggesting that they would die
as H-free WR stars and possibly GRBs.  \citet{eldridge11} therefore
inferred that there may be a population of relatively isolated WR
stars and GRBs.  In this paper we have argued something different:
that LBVs might be the direct products of mass transfer in binaries,
and in particular, the mass gainers. (If this is correct, it may
suggest that quasi-homogeneous chemical evolution may not be
appropriate for these stars, but a further discussion is beyond the
scope of this paper.)  We have discussed above that this binary
scenario alleviates the apparent paradox of LBVs exploding as SNe~IIn,
which is prohibited in the standard single-star evolution framework
(although see \citealt{groh13} for a possible caveat).  If LBVs do
indeed explode as SNe~IIn, then we should expect the relative
isolation of LBVs discussed herein to be reflected in their resulting
SNe as well.  As noted above, if these stars receive a kick when a
companion in the binary system explodes, they may travel 100 pc or
more from their birthsites by the time they die.  They should die at
locations that are preferentially outside clusters, causing them to
systematically avoid bright H~{\sc ii} regions or blue star clusters.
Even without a kick, the extended lifetime of an LBV (because its true
initial mass was much lower than its apparent $M_{\rm eff}$) would
allow it to live much longer then the H~{\sc ii} region that it might
have been born in, since the massive stars that ionized the region may
have long since died away.

Evidence consistent with this from observations of SN host
environments has already been published in the literature, although
the results were interpreted somewhat differently.  As noted in the
introduction, Anderson and collaborators
\citep{aj09,anderson12,habergham14} have found that SN~IIn show a
weaker association with H$\alpha$ emission in host galaxies, as
compared to SNe~Ibc, and their H$\alpha$ association is more like that
of SNe~II-P.  Those authors interpreted these results in a context
where initial mass dominates SN types; they attributed the correlation
of SNe~Ibc with H$\alpha$ to indicate the highest range of initial
mass, and they interpreted the lack of correlation between SNe~IIn and
H$\alpha$ as indicating lower initial masses.  They therefore
concluded that LBVs are a less likely progenitor channel for SNe~IIn.

As we noted earlier, this result motivated us to examine the
environments of nearby LBVs.  We found that despite their high
luminosities and initial masses higher than those of most RSGs, LBVs
do actually appear to be very isolated from OB star clusters (moreso
than WR stars, and almost as isolated as RSGs).  We have argued that
this is most likely due to receiving a kick from a companion's SN that
preferentially sends the mass gainers flying out of clusters.  Binary
evolution will have an impact not only on the stars that are kicked
(in this case the LBVs and resulting SNe~IIn), but also on the mass
donors that explode first.  They (in this case SNe~Ibc or IIb) will
occur preferentially in young clusters as compared to other SN types
that are found anywhere, and will be anticorrelated with those that
are preferentially kicked out of clusters.  Thus, assuming that both
SNe~Ibc and SNe~IIn come from a wide range of overlapping initial
masses appears to be consistent with the data.  Because of the
important influence of binary evolution, interpreting SN host
invironments only in terms of progenitor initial mass is overly
simple.\footnote{Interestingly, association with clusters does have
  something to do with initial masses, but not in the way that it is
  usually discussed.  In the binary scenario, a SN located in a
  cluster favors the star that is the initially {\it more massive of
    the two stars} and explodes first, rather than selecting stars
  preferentially above some particular mass value.}  Some stars and
certain types of SNe can selectively move from their birthsites; a
theoretical study of the delay-time distributions and correlations (or
not) with H~{\sc ii} regions is needed to quantify this effect.

Moreover, there are reasons why a correlation with H$\alpha$ as
adopted by Anderson and collaborators is not necessarily a reliable
indicator of extreme youth and the highest intial masses.  That
interpretation was criticized by \citet{crowther13}, who pointed out
that the brightest (and therefore the most easily detected H~{\sc ii}
regions in distant galaxies) are giant H~{\sc ii} regions that are
actually quite long lived ($\sim$20 Myr), with ages comparable to the
ages of stars with initial masses of only $\sim$12~$M_{\odot}$.
Anderson and collaborators have based their metric of association with
H$\alpha$, and hence youth, as flux-depenent, so that a SN landing on
a bright source of H$\alpha$ is interpreted as younger than a SN
landing on pixels with fainter H$\alpha$.  In some sense this is
backwards --- the brightest H~{\sc ii} region complexes are the
longest lived ones with median ages $>$10 Myr, whereas more isolated,
smaller, and fainter H~{\sc ii} regions are fleeting, and have shorter
ages of only $\sim$3-4 Myr.  Therefore, finding a SN in a smaller
isolated H~{\sc ii} region is more likely to indicate a very high
initial mass above 60 $M_{\odot}$ than if a SN is seen in a very
bright and long-lived giant H~{\sc ii} region complex
\citep{crowther13}. When these fainter and more isolated H~{\sc ii}
regions go preferentially undetected in ground-based imaging of
distant galaxies, their associated SNe may be mistakenly assumed to
come from an older population.  Note that SN~1987A, with an initial
mass of $\sim$18 $M_{\odot}$ would have a high correlation with star
formation as seen from a distant galaxy because of its proximity to 30
Dor (only $\sim$1{\arcsec} at a distance of 50 Mpc), whereas $\eta$
Carinae would explode in the fainter Carina Nebula, which in some
cases would not be detected.

Anderson et al.\ have discussed potential biases in detecting various
SN types and found that it does not strongly impact the trends in
their observations, but there may be other selection effects besides
those that govern the discovery of various SN types.  For example, a
metallicity bias may cause there to be more of some type of SN in the
inner regions of a galaxy as compared to outer regions.  In a typical
spiral galaxy, inner regions are more densely packed with star
formation, and they have a larger fraction of the observed surface
area covered by very bright H~{\sc ii} regions than in the much
sparser outer regions.  Indeed, \citet{habergham14} found that SNe~Ibc
in their sample were systematically found at smaller galactic radii
than SNe~II.  Thus, even if SN types have the same range of progenitor
initial mass, a type of SN that has a preference for smaller
galactocentric radii because of metallicity may have a higher probably
of coincidence with a very bright H~{\sc ii} region.  The potential
influence of this bias deserves further attention.
%Also, fainter isolated H~{\sc ii} regions that might favor very young
%and massive stars will more easily escape detection in ground-based
%images.

The sensitivity and potential bias in the flux-dependent H$\alpha$
diagnostic used by Anderson and collaborators
\citep{aj09,anderson12,habergham14} may explain why their conclusions
differed from other studies of SN host environments. Earlier studies
of spatial proximity to H~{\sc ii} regions (rather than the strength
of H$\alpha$ flux) found no statistically significant preference
between SNe~II and Ibc \citep{vandyk92,vandyk96,bartunov94}.  Some of
these might be chance coincidence with long-lived giant H~{\sc ii}
regions at relatively course ground-based angular resolution.  In
general, one expects no direct correlation between a SN and its own
H~{\sc ii} region for any but the most massive stars above
75~$M_{\odot}$ \citep{crowther13}.  Indeed, \citet{smartt+09} found
that high resolution {\it HST} images for nearby SNe with good
constraints on the progenitor stars do not support a correlation with
H~{\sc ii} regions, and \citet{smartt09} argued against the monotonic
increasing progenitor mass from Type~II to IIb to Ibc based on this
and a number of other considerations (see also \citealt{crowther13}).
When examining the SN host color at the location of SNe rather than
host H$\alpha$, \citet{kelly12} found that SNe~Ibc, SNe~II, and
SNe~IIn show basically the same distribution of
$u^{\prime}-z^{\prime}$ colors.  They did, however, find bluer colors
for broad-lined SNe~Ic and SNe~IIb; it is unlikely that this is due to
initial mass, since progenitor detections suggest rather low masses
for SNe~IIb.

One may ask, if LBVs are the mass gainers that get a kick from their
companion's SN~Ibc in a cluster or association, then why do we not
detect bright LBVs at the positions of SNe~Ibc?  The most likely
answer is that the mass gainer will appear as a more luminous but hot
O-type star that is still quite difficult to detect (see
Figure~\ref{fig:hrd}), and may not become a visually bright LBV until
1-2 Myr later, when its own core evolution drives it off the main
sequence to become a cooler B[e]/LBV supergiant.  Moreover, in the
binary scenario, the majority of SNe~Ibc will be from relatively low
initial masses, compared to the high-mass WR progenitors envisioned in
the single-star scenario.

Overall, we find it likely that the sequential order of explosion in a
binary system and subsequent kicks out of clusters (and to a somewhat
lesser extent, metallicity and star cluster density), plus the
extended main sequence lifetime before mass accretion, will have a
more important influence on the observed environments of core-collapse
SNe than monotonic differences in the range of initial mass.  This may
be especially true for more extreme or rare explosions that may
require special evolutionary paths in binaries, such as long GRBs
\citep{kelly14,fruchter06}.

Of course, the observed isolation of SNe~IIn might also be influenced
to some degree by SN~IIn progenitor contamination from non-LBV
objects.  Indeed, there is strong evidence for this already, although
the fraction of contamination is poorly constrained.  In principle,
any type of explosion can yield a Type~IIn event, since the IIn
designation depends on CSM interaction and not the explosion
mechanism.  While LBV mass loss fits the bill \citep{smith14}, the
case for LBVs is strongest for the most luminous SNe IIn where the CSM
mass budget requires extreme parameters \citep{smith14}.  B[e]
supergiants are also suitable for providing moderate-luminosity
SNe~IIn, and their spatial distribution is even more isolated than
LBVs, comparable to RSGs (Figures \ref{fig:cum} and \ref{fig:cum2}).
Moreover, the concentration of eruptive mass loss associated with the
final nuclear burning sequences in the last few years before core
collapse \citep{sa14,qs12} means that the progenitor need not have
been in a dense-wind phase for very long.  Besides massive LBVs
exploding as SNe~IIn, there are also examples where SNe~IIn may result
from: (1) extreme RSGs with strong clumpy winds \citep{shr09,smith09},
(2) electron-capture SNe (ecSNe) from 8-10 $M_{\odot}$ super-AGB stars
(such as the class of SNe~IIn-P and possibly the Crab Nebula;
\citealt{mauerhan13,smith13,chugai04a}), and (3) some hybrid Type
Ia/IIn events that are apparently thermonuclear SNe~Ia exploding in a
dense H-rich CSM, like SN~2002ic and several similar objects
\citep{chugai04,silverman13}.  When these other objects having much
lower mass progenitors are included with SNe~IIn, it would make the
mixed population appear statistically even more isolated.  This is the
likely explanation for why \citet{habergham14} found that SNe~IIn had
a spatial distribution similar to SNe~II-P, even though we find LBVs
have isolation intermediate between WR stars and RSGs.

It is worth reiterating the point that not all mass gainers will
become LBVs, since this likely depends on the mass accreted and final
luminosity (many may become B[e] supergiants instead, or they may
evolve to become RSGs, etc.), and not all mass gainers will
necessarily become SNe~IIn, since these SNe require special conditions
(pre-SN mass ejections).  Comparing the observed fraction of SNe IIn
(8--9\% of core collapse SNe; \citealt{smith11}) to theoretical
expectations requires continued modeling of the evolution of mass
gainers, which still depends on uncertain assumptions.

In addition to core-collapse SNe, caution about inferring an initial
mass also applies to SN impostors and related transients.  Some
eruptive transients have been discussed where the apparent age of
surrounding stellar population indicates a surprisingly low initial
mass.  The most representative cases are SN~2008S and the 2008
transient in NGC~300 \citep{thompson09,prieto08,gogarten09}.  If LBVs
or lower-mass analogs preferentially receive a kick from an exploded
companion, then this needs to be included in the uncertainty when
inferring an age and initial mass from their surrounding stellar
populations. One last important consequence to mention is that stars
with initial masses below 8 $M_{\odot}$ may, if in a binary system,
accrete mass and explode as a ccSN or an electron-capture SN, even if
their initial mass would have precluded that fate in a single-star
scenario.  Due to the slope of the IMF, these initially lower-mass
stars may potentially contribute a substantial number of observed SNe.
We encourage a detailed study of this effect, including the delay time
distribution and its contribution to SN statistics.

\section{EPILOGUE}

This paper suggests a major reversal in our interpretation of the
nature of LBVs.  Instead of being the result of instability in very
massive single stars that have evolved to become unstable through
their own core evolution and mass loss, they must preferentially or
exclusively be the products of binary evolution. They are most likely
the mass gainers in binary RLOF or mergers because their ages and
initial masses (inferred from single-star evolution tracks)
systematically disgree with the stars around them.  While counter
arguments can point to one or two possible exceptions, this must apply
to most of the LBVs.

Why was this seemingly obvious result missed?  The answer may be
partly sociological: LBVs are very few in number, and $\eta$ Car
garners disproportionate attention, so its location in the Carina
Nebula did not arouse suspicion.  Also, interest in LBVs has
concentrated more on driving their eruptive mass loss, as opposed to
the evolutionary history that led them to be unstable in the first
place.  Last, LBVs seem to fit nicely into the picture of single-star
evolution (and indeed they are needed for it), so the question of
whether or not their environments contradict models was not pursued
with adequate vigor.

Actually, a suggestive result along similar lines was found more than
a decade ago by N.L.\ King and collaborators (2000; see also
\citealt{king97,king98}, as well as \citealt{kg85}).  Based on
environments of some candidate LBVs in M31, it was found that they
seem to reside outside OB associations and clusters, implying stellar
ages of several Myr.  That work did not instigate a major rethinking
of our standard paradigm of massive star evolution because the paper
was not accepted for publication in the refereed literature (the first
author has since left the field).  In hindsight, the main result of
that work appears to have pointed in the correct direction after all.

%\acknowledgments \footnotesize
\smallskip\smallskip\smallskip\smallskip
\noindent {\bf ACKNOWLEDGMENTS}
\smallskip
\footnotesize

NS thanks J.S.\ Gallagher for discussions regarding binary evolution
and LBVs, and in particular for drawing our attention to previous work
by N.\ King.  We benefitted from discussion about SN host environments
with P.\ Crowther, J.\ Anderson, and P.\ Kelly, and from further
discussions concerning binary evolution with S.\ de Mink. We also
thank the referee (J. Eldridge) for a careful reading and thoughtful
comments that improved the paper.  Partial support was provided by NSF
award AST-1312221.  This research has made use of the SIMBAD database,
operated at CDS, Strasbourg, France.

%{\it Facilities:} LBT (MODS); MMT (Bluechannel)

% REFERENCES

\end{document}